# Single-stripe magnetic order and bilayer spin dynamics in single-crystalline $La_3Ni_2O_7$

Lixing Chen[1#], Enkang Zhang[1#], Yiqing Hao[2], Yinghao Zhu[1], Bingkun Cui[1], Douglas L. Abernathy[2], Travis J. Williams[3], Yoichi Ikeda[4], Hao Zhang[1], Feiyang Liu[1], Wenbin Wang[5], Qisi Wang[6] & Jun Zhao[1,5,7]*

[1]*State Key Laboratory of Surface Physics and Department of Physics, Fudan University, Shanghai 200433, China*

[2]*Neutron Scattering Division, Oak Ridge National Laboratory, Oak Ridge, Tennessee 37831, USA*

[3]*ISIS Neutron and Muon Source, STFC Rutherford Appleton Laboratory, Didcot, Oxfordshire OX11 0QX, United Kingdom*

[4]*Institute for Materials Research, Tohoku University, Katahira, Aoba-ku, Sendai 980-8577, Japan*

[5]*Institute of Nanoelectronics and Quantum Computing, Fudan University, Shanghai 200433, China*

[6]*Department of Physics, The Chinese University of Hong Kong, Shatin, Hong Kong, China*

[7]*Shanghai Research Center for Quantum Sciences, Shanghai 201315, China*

## Abstract

**High-temperature superconductivity in pressurized and thin-film nickelates has generated intense interest, yet magnetism in their ambient-pressure parent phases remains poorly understood, despite its potentially crucial role in pairing. Progress has been limited by the lack of large single crystals suitable for neutron spectroscopy. Here we overcome this limitation and map the spin order and excitations of single-crystalline $La_3Ni_2O_7$. We find single-stripe magnetic order with a finite spin gap, anisotropic in-plane excitations and pronounced bilayer modulation revealing antiferromagnetic interlayer coupling. The dispersion is captured by a bilayer Heisenberg-type model with strong interlayer exchange and competing in-plane couplings. Absolute normalization reveals that, although the spin-wave bandwidth is only about 25% of that in cuprates, the local dynamic susceptibility at comparable energies is enhanced, yielding a comparable total fluctuating moment. These results highlight intense mid-energy spin excitations rooted in substantial electronic correlations, establishing a magnetic framework distinct from cuprates and relevant to superconductivity in nickelates.**

Since the discovery of high-temperature cuprate superconductors nearly four decades ago, a central challenge has been to identify the interactions that mediate electron pairing. Because high-temperature superconductivity develops when static antiferromagnetic order in the parent compound is suppressed by carrier doping, magnetism is widely believed to play a pivotal role in the pairing mechanism[1–3]. However, despite decades of intensive research, the microscopic interactions that control high-temperature superconductivity have not yet been established conclusively.

High-temperature superconductivity in pressurized Ruddlesden–Popper (RP) nickelates[4,5] and in epitaxially strained nickelate thin films[6–8] offers a fresh perspective on unconventional pairing. Although nickelates have similar layered structural motifs as cuprates, they differ in their electronic configurations and orbital hierarchy[9]. In cuprates, the low-energy electronic structure is dominated by a single Cu $3d_{x^2-y^2}$ orbital, forming a nearly ideal two-dimensional antiferromagnetic state[1]. By contrast, Ruddlesden-Popper nickelates host mixed-valence Ni ions with $3d^7$ and $3d^8$ configurations, where both $d_{x^2-y^2}$ and $d_{z^2}$ orbitals can contribute near the Fermi level[4,5]. The $d_{z^2}$ orbital couples through apical oxygens, mediating interlayer correlation, whereas the $d_{x^2-y^2}$ orbital primarily governs in-plane correlations, as for cuprates. The participation of several orbitals can generate orbital-selective degrees of freedom, alter the underlying magnetic exchange pathways[10–16], and potentially lead to qualitatively different pairing interactions[17–30]. Such similarities and differences place nickelates at a unique crossroads between cuprates and other correlated superconductors, offering an unprecedented opportunity to disentangle the intertwined roles of spin fluctuations and orbital character in high-critical temperature ($T_c$) superconductivity.

In particular, bilayer $La_3Ni_2O_7$ exhibits superconductivity under pressure or compressive epitaxial strain with transition temperatures reaching around 90 K, placing it among the highest $T_c$ unconventional superconductors outside the cuprates[4,7,8,31]. At ambient conditions, $La_3Ni_2O_7$ develops antiferromagnetic order, establishing it as the parent compound of the pressurized or thin-film superconducting phases. Previous studies of resonant X-ray scattering on single crystals or thin films reported a magnetic Bragg peak at (0, 0.5, $L$) in the orthorhombic notation, corresponding to (0.25, 0.25, $L$) in the pseudotetragonal notation (Supplementary Fig. S9), consistent with either stripe or double-stripe order[32–35] . This is consistent with muon spin rotation and nuclear magnetic resonance measurements, which likewise detect static magnetism[36–38]. A recent synchrotron X-ray diffraction study further reported charge order with inequivalent Ni sites in $La_3Ni_2O_7$, highlighting the importance of charge disproportionation in this material[39] . Elastic neutron diffraction on powders identified a more complex situation: although some studies reported no magnetic order[40,41], others found two coexisting magnetic wavevectors, $\boldsymbol{q}$ = (0, 0.5, 0) and (0, 0.5, 0.5), suggesting possible phase separation[42]. Dynamically, resonant inelastic X-ray scattering from single crystals revealed spin excitations that are compatible with either stripe or double-stripe scenarios[32], and the excitations in superconducting films under compressive strain[35] are rather like those in the bulk. This suggests that spin correlations relevant to pairing are already present in the parent phase, as for cuprates[43]. On the other hand, inelastic neutron scattering from polycrystalline samples identified only a weak flat spin-fluctuation signal near 45 meV (ref. [40]). Both the resonant inelastic X-ray and neutron scattering results have been interpreted in terms of strong interlayer magnetic interactions. However, a complete picture of the momentum- and energy-resolved spin correlations across the entire Brillouin zone, including the low-energy spin gap, in-plane anisotropy, out-of-plane modulation, absolute local dynamic susceptibility, total fluctuating magnetic moment and the precise magnetic interactions and structure, remains to be established. Such information is essential for establishing the role of electron correlation and for constraining microscopic models of magnetism and superconductivity.

Progress in this field has long been hindered by the formidable challenge of synthesizing large $La_3Ni_2O_7$ single crystals suitable for advanced spectroscopic probes, such as neutron scattering. We overcome this bottleneck by optimizing high-pressure optical floating-zone growth, enabling the production of millimeter- to centimeter-sized crystals. This advance enabled us to map the magnetic correlations of $La_3Ni_2O_7$ over broad ranges of momentum and energy. Our measurements established single-stripe magnetic order, revealed strong antiferromagnetic coupling between the two $NiO_2$ layers of a bilayer and quantified both the competing in-plane interactions and the absolute spectral weight of the spin fluctuations. These results establish a quantitative magnetic framework for the ambient-pressure parent phase of superconducting bilayer nickelates.

We first characterized the $La_3Ni_2O_7$ single crystals. Figure 1c shows a representative magnetic susceptibility curve with a clear anomaly near 151 K, marking the onset of a magnetic phase transition. Moreover, crystals from the same growth batch have been shown to exhibit robust superconductivity under pressure, characterized by sharp superconducting transitions, zero resistance and well-defined superconducting gaps[44].

To investigate the magnetic ordering, we performed neutron diffraction measurements. In contrast to previous studies on powder samples that reported either two sets of magnetic ordering peaks or no magnetic peaks at all, our single crystals display well-defined magnetic Bragg peaks corresponding to $\boldsymbol{q}$ = (0, 0.5, 0.5), with no evidence for peaks at $\boldsymbol{q}$= (0, 0.5, 0) (Fig. 1d and Supplementary Fig. S2). This confirms the presence of a single magnetic phase without phase separation, which provides an ideal foundation for determining the precise magnetic structure and exchange interactions. The intensity of the (0, 0.5, 2.5) peak as a function of temperature indicates the onset of magnetic order at 151 K and aligns with the susceptibility measurements (Fig. 1c). The elastic neutron diffraction intensity observed at $\boldsymbol{Q}$ = (0, 0.5, 0.5) and equivalent positions can, in principle, be explained by either a stripe- or a double-stripe-type magnetic structure, both characterized by antiferromagnetic alignment between adjacent bilayers, as the two models yield similar magnetic structure factors (Fig. 1a, b and Supplementary Fig. S4a).

To further resolve the magnetic ground state and interactions, we carried out inelastic neutron scattering measurements. Figure 2 displays constant-energy maps of the spin excitations in the ($H$, $K$, 2.5) plane of reciprocal space, acquired at 6 K, well below the magnetic ordering temperature. The excitations are centred at $\boldsymbol{Q}$ = (0.5 ± $n$, 0 ± $m$, 2.5) and $\boldsymbol{Q}$ = (0 ± $n$, 0.5 ± $m$, 2.5), where $n$ and $m$ are integers. Although the underlying lattice symmetry is orthorhombic with the space group *Amam*, the presence of 90° twin domains, arising from the nearly equivalent lattice constants $a$ and $b$, leads to a four-fold-like symmetry of the magnetic excitations. Energy cuts through the excitation signal revealed a spin gap of about 5 meV (Fig. 1f), consistent with the constant-energy maps (Fig. 2a, b). The finite gap reflects an appreciable magnetic anisotropy, consistent with the easy-axis orientation along the *a* direction. It therefore provides another low-energy scale that constrains the spin dynamics beyond the dominant exchange interactions.

With increasing energy (Fig. 2a–j), the excitations at (0, 0.5, 2.5) disperse outwards anisotropically, with a steeper dispersion along the *K* direction than along *H*, forming an elliptical ring at ~35 meV. Along the *H* direction, the excitations merge with those from the neighboring Brillouin zone at around 49 meV, producing a square-like pattern (Fig. 2f,g). At higher energies, the signal forms a ring at (0, 2, 2.5) and eventually reaches the band top near 70 meV (Fig. 2h-j).

To better visualize the dispersion in energy–momentum (*E*–***Q***) space, we projected the excitations along high-symmetry paths. As shown in Fig. 3a, the dispersion along (0, 1.3, 2.5) → (0, 2, 2.5) corresponds to the steeper branch, whereas the path (-0.5, 1.5, 2.5) → (0, 1.5, 2.5) follows the flatter branch and highlights pronounced zone-boundary softening. The combination of softening and anisotropic dispersion is indicative of underlying competing magnetic interactions, as we discuss later. Another branch near (0, 1.5, 2.5) was resolved with a gap of around 25 meV.

To quantify the dispersions and intensities of the spin excitations, we performed constant-energy cuts at various energies. As shown in Fig. 4a–g, the single peak centered at (0, 0.5, 2.5) at 6 meV evolves into a pair of peaks along the *K* direction for $E \geq 20$ meV. Similar behavior was observed for cuts along the *H* direction, albeit with a more gradual dispersion, consistent with the anisotropy revealed in the constant-energy images and $E$–$\boldsymbol{Q}$ plots (Figs. 2 and 3).

Having established the in-plane momentum dependence of the spin excitations, we next examined their out-of-plane behavior by performing cuts along the *L* direction. The excitation intensity exhibited pronounced *L*-dependent modulations, consistent with bilayer periodicity and providing direct evidence for antiferromagnetic interlayer spin correlations (Fig. 1e, Supplementary Fig. S7). The spectral weight was dominated by the acoustic branch, which showed a sine-like modulation and reached its maximum near $L \approx 2.6$ (Fig. 1e, Supplementary Fig. S7), whereas the optical branch remained much less intense and was resolved as a feature at the complementary position near $L \approx 5.2$ (Supplementary Fig. S8).

To quantitatively elucidate the magnetic interactions underlying the gapped, anisotropic in-plane spin-excitation dispersion together with the interlayer correlations in $La_3Ni_2O_7$, we carried out a systematic analysis within the framework of a Heisenberg-type Hamiltonian (Fig. 1a, b). The model Hamiltonian can be expressed as:

$$H = \sum_i J_c \boldsymbol{S}_i^t \cdot \boldsymbol{S}_i^b + \sum_{\langle i,j \rangle} J_2 \boldsymbol{S}_i^\alpha \cdot \boldsymbol{S}_j^\alpha + \sum_{\langle i,j \rangle_a} J_{1a} \boldsymbol{S}_i^\alpha \cdot \boldsymbol{S}_j^\alpha + \sum_{\langle i,j \rangle_b} J_{1b} \boldsymbol{S}_i^\alpha \cdot \boldsymbol{S}_j^\alpha + \sum_i A(\boldsymbol{S}_i \cdot \boldsymbol{n})^2$$

Here $\alpha$ denotes the layer index for the top (t) or bottom (b) layer. $\boldsymbol{S}_i^\alpha$ is the spin operator at magnetic Ni site $i$ in layer $\alpha$. $J_c$ is the interlayer exchange interaction, and $J_2$ is the next-nearest-neighbor exchange coupling. $J_{1a}$ and $J_{1b}$ represent the nearest-neighbor exchange couplings along the $a$ and $b$ directions, respectively. $A$ denotes the single-ion anisotropy and $\boldsymbol{n}$ denotes a unit vector along the $a$ axis (Fig. 1a and b). This minimal Hamiltonian does not include an explicit coupling between neighboring bilayers. Although the half-integer $L$ dependence of the elastic magnetic peaks shows that adjacent bilayers are antiferromagnetically phased, the energy scale selecting this inter-bilayer phase relation is expected to be much smaller than the dominant intra-bilayer and in-plane interactions. The microscopic origin of this weak phasing interaction is beyond the scope of the present minimal model and may involve weak residual inter-bilayer exchange or subtle lattice distortions that lift the geometrical frustration.

Using the SpinW package, we simulated with linear spin-wave theory the spin-excitation spectra along high-symmetry directions, systematically varying the interaction parameters to capture the experimental dispersions. As an initial attempt, we employed a model containing only in-plane exchange interactions (Supplementary Fig. S6). This model reproduced a low-energy branch near (0, 0.5, 2.5) but incorrectly predicted further low-energy intensity at (0, 1, 2.5), inconsistent with its experimental identification as a zone boundary. The pronounced $L$-dependent intensity modulation with bilayer periodicity indicates the necessity of including substantial antiferromagnetic interlayer coupling. Indeed, only by introducing a sizeable interlayer exchange $J_c$ could these discrepancies be resolved, underscoring the critical role of interlayer correlations in $La_3Ni_2O_7$.

Because the previously proposed single-stripe and double-stripe spin configurations yield nearly identical magnetic wavevectors and static structure factors, it is difficult to unambiguously differentiate them with elastic scattering alone. Inelastic neutron scattering, with momentum- and energy-resolved spectra over the entire Brillouin zone, provides a sensitive fingerprint of the underlying spin configuration. In comparative spin-wave simulations, the double-stripe model yielded an almost isotropic in-plane dispersion, in clear disagreement with the data (Fig. 2a-j and 3a and Supplementary Fig. S4). Introducing in-plane anisotropic exchanges or further neighbor couplings in the double-stripe model induced some dispersion anisotropy, but the resulting intensity distribution contradicted experimental observations (Fig. 2a-j and Supplementary Fig. S5c-j). Moreover, this model also predicted further low-energy branches near the zone center that are not observed (Fig. 3a and Supplementary Fig. S5b). By contrast, the single-stripe configuration naturally reproduced the observed characteristics: the pronounced in-plane anisotropy and the softening of the zone-boundary excitation near (0.5, 0.5, 2.5), both originating from competing magnetic interactions between intra- and interstripe antiferromagnetic couplings. With optimized exchange coupling

parameters $SJ_c$=39.86 ± 0.85, $SJ_{1a}$=2.36 ± 0.22, $SJ_{1b}$=3.71 ± 0.44, $SJ_2$=4.63 ± 0.18 and $SA$=-0.07 ± 0.015 meV, the calculated dispersions quantitatively reproduced the measured spectra. Here, S denotes the spin magnitude, and the quoted uncertainties represent the statistical uncertainties within the adopted stripe-type linear spin-wave fit. Possible systematic effects associated with background subtraction and model choice are not included. The simulations captured not only the overall bandwidth and energy scale but also the subtle momentum dependence along high-symmetry directions (Figs. 3a,b). Moreover, the computed constant-energy maps reproduced the anisotropic elliptical scattering features observed experimentally (Fig. 2). Within the present spin-wave framework, the fitted parameters are tightly constrained by these momentum- and energy-resolved features and provide a consistent description of the dominant magnetic interactions.

In the effective spin-wave description, the stripe spin order, consisting of alternating magnetic and non-magnetic stripes and breaking the in-plane $C_4$ symmetry, can be stabilized by the relatively strong next-nearest-neighbor antiferromagnetic coupling $J_2$ along the diagonal direction (Fig. 1b). The nearest-neighbor antiferromagnetic exchange $J_{1b}$ promotes antiferromagnetic alignment between adjacent spin stripes, whereas the weaker antiferromagnetic coupling $J_{1a}$ between parallel-aligned spins within a stripe introduces magnetic frustration. This frustration or competing interaction gives rise to the observed softening of the transverse spin-wave branch and the pronounced in-plane anisotropy. The strong interlayer exchange $J_c$ originates from superexchange interactions mediated by apical oxygens associated with the $d_{z^2}$ orbitals between adjacent $NiO_2$ layers, where the Ni–Ni distance is relatively short. This exchange hierarchy contrasts with that of bilayer cuprates. In bilayer $YBa_2Cu_3O_{6+x}$, acoustic and optical spin waves are described by an in-plane exchange $J_{ab} \approx 125$ meV and an intrabilayer exchange $J_c \approx 11$ meV, so the bilayer coupling is much weaker than the in-plane exchange[45–47] . In $La_3Ni_2O_7$, by contrast, $SJ_c$ is one of the largest magnetic energy scales, larger than the effective in-plane couplings, revealing a reversed bilayer-to-in-plane exchange hierarchy.

Analysis of the elastic neutron diffraction data yielded an ordered magnetic moment of 0.67(3) $\mu_B$ per ordered Ni ion in the single-stripe model, where $\mu_B$ is the Bohr magneton, which is substantially smaller than that expected in the ionic limit for a fully localized $S = 1$ system, consistent with the presence of competing magnetic interactions and, possibly, also with the influence of Ni-O covalency and electronic itinerancy. These results suggest that a bilayer Heisenberg model with a delicately balanced interplay of strong interlayer coupling and competing in-plane interactions, together with a single-stripe spin configuration, may provide a consistent description of the spin order and dynamics in $La_3Ni_2O_7$.

Further insight into the nature of magnetic correlations and their implications for superconductivity could be obtained from the energy dependence of local dynamic susceptibility, transformed into absolute units by normalization to the acoustic phonons.

This procedure enables a direct, quantitative comparison with cuprates, other superconductors and theoretical models. As shown in Fig. 5, in addition to a spin gap below ~ 5 meV, the absolute local dynamic susceptibility exhibits a continuum of mid-energy spin excitations within 10–70 meV. By integrating the magnetic signal over energy and including the elastic ordered contribution, we obtained a total magnetic spectral weight of 1.75±0.25 $\mu_B^2$ per ordered $Ni^{2+}$ on the stripe, corresponding to 0.88 ±0.13 $\mu_B^2$ per total planar Ni site, comparable with that of cuprates. As electronic itinerancy typically suppresses the dynamic magnetic moment, the persistence of a considerable fluctuating moment indicates that itinerant electrons coexist with local magnetism in $La_3Ni_2O_7$, sustained by substantial electron correlations. For context, the insulating single-layer nickelate $La_{2-x}Sr_xNiO_4$ also exhibits well-defined spin waves[48,49] . The key distinction in bilayer $La_3Ni_2O_7$ is that stripe-type in-plane correlations coexist with strong intrabilayer antiferromagnetic coupling and acoustic and optical spin-wave branches, revealing a bilayer magnetic framework absent in the single-layer nickelates.

Although the total fluctuating moments of both systems are comparable, the overall spin-excitation bandwidth in $La_3Ni_2O_7$ is limited to ~70 meV, far smaller than the ~300 meV bandwidth of cuprates. Because the change in bandwidth is rather small between the bulk $La_3Ni_2O_7$ and the superconducting phase under compressive strain[32,35], the fluctuation moment and overall spin-excitation spectra are expected to remain comparable between the parent and superconducting states, in accordance with the magnetic moment sum rule. Yet the superconducting critical temperatures of nickelates and cuprates are of similar magnitudes. This contrast indicates that a large spin-excitation bandwidth is not, by itself, the decisive ingredient for pairing. Rather, the density and character of spin fluctuations near the relevant electronic energy scales seem to be key. Indeed, the absolute local dynamic susceptibility in the 10 meV to 70 meV range is markedly larger in $La_3Ni_2O_7$ than in cuprates (Fig. 5), suggesting that these dense mid-energy spin excitations couple more effectively to electrons on the energy scale relevant to Cooper pairing.

In $La_3Ni_2O_7$, these relevant energy scales may arise from two complementary channels. The first is the interlayer coupling between adjacent Ni ions, mediated by apical oxygens and dominated by the $d_{z^2}$ orbital. This coupling may directly produce excitations with energies of the same order as the superconducting gap[44,50–52]. Second, intralayer exchange interactions within the $NiO_2$ planes, although individually modest, combine to produce collective excitation energies up to about 30-40 meV, again comparable with the gap[44,50–52] (Supplementary Fig. S6). These excitations are mainly linked to the $d_{x^2-y^2}$ orbital and provide another reservoir of spectral weight. Such a unique combination of interlayer and intralayer interactions creates a high density of mid-energy spin excitations that are well matched to the pairing scale.

The quantitative determination of the magnetic interactions in $La_3Ni_2O_7$ provides important insight into the pressure- and strain-tuned magnetic phase diagram, as well as its relation to superconductivity. Within a local-moment framework, the in-plane magnetic interactions are sufficient to stabilize the single-stripe magnetic order. Moreover, the combination of sizeable interlayer coupling and competing in-plane exchange interactions places this magnetic state in a delicate balance, making it sensitive to external perturbations, such as pressure and in-plane strain. Under pressure or compressive epitaxial strain, $La_3Ni_2O_7$ is expected to undergo subtle structural changes, including a reduction of the in-plane orthorhombicity and a suppression of $NiO_6$ octahedral tilts. Such a higher-symmetry structure could reduce the lattice pinning of the stripe-type order and weaken the static magnetic order, while allowing strong dynamic spin correlations to persist. This suggests a scenario in which superconductivity emerges not from the loss of magnetic interactions but from a redistribution of the magnetic spectral weight from static order to fluctuating correlations. Within the effective spin-wave description, these structural changes can also modify the balance among $J_{1a}$, $J_{1b}$, $J_2$ and $J_c$, providing a magnetic route by which pressure or strain tunes the competing interactions identified here. The involvement of both $d_{x^2-y^2}$ and $d_{z^2}$ orbitals further suggests that superconductivity in bilayer nickelates is shaped by the cooperation of in-plane and interlayer magnetic correlations. The ability to grow large $La_3Ni_2O_7$ single crystals and determine their spin dynamics in absolute units provides a quantitative materials benchmark for understanding how structural tuning, orbital selectivity and magnetic spectral weight are linked in superconducting bilayer nickelates.

This work establishes a quantitative magnetic framework for $La_3Ni_2O_7$: a stripe-type magnetic ground state with competing in-plane interactions and strong antiferromagnetic interlayer coupling, effectively captured by a bilayer Heisenberg-type model. Although its spin-wave bandwidth is only about 25% of that in cuprates, $La_3Ni_2O_7$ exhibits an enhanced mid-energy local dynamic susceptibility and a comparable total fluctuating moment, reflecting robust electronic correlations and magnetic interactions persisting into its metallic state. The presence of intense mid-energy spin excitations, coupled with bilayer magnetism, outlines a new magnetic framework distinct from that of cuprates. Our findings suggest that the key to superconductivity in bilayer nickelates may not lie in achieving large spin-excitation bandwidths but rather in generating a dense spectrum of magnetic fluctuations concentrated near the pairing energy scale. The cooperative involvement of both $d_{x^2-y^2}$ and $d_{z^2}$ orbitals, coupled through interlayer and intralayer exchange pathways, naturally provides such an environment, linking orbital selectivity and spin dynamics to the emergence of high-temperature superconductivity. By quantitatively establishing the absolute scale and character of spin order and dynamics in single-crystalline $La_3Ni_2O_7$, this work places stringent experimental constraints on theoretical models and provides a foundation for understanding pressure- and strain-induced superconductivity across this class of materials.

[#] These authors contribute equally to this work.

[*]Correspondence and requests for materials should be addressed to J.Z.

(zhaoj@fudan.edu.cn)

**Acknowledgements** We thank Manabu Okawara and Gen Takahashi for technical assistance at JRR-3. A portion of this research used resources at the Spallation Neutron Source, a DOE Office of Science User Facility operated by the Oak Ridge National Laboratory. The neutron diffraction experiments were performed under the Joint-Use Research Program for Neutron Scattering, Institute for Solid State Physics, University of Tokyo, at JRR-3 (6G IRT program 25402), and the GIMRT research program of the Institute for Materials Research, Tohoku University (202508-QBKNE-0507). Part of the measurements was performed using the facilities of the Fudan Nano-fabrication Laboratory.

**Funding** This work was supported by the Key Program of the National Natural Science Foundation of China (grant no. 12234006), the Shanghai Municipal Science and Technology Project (grant no. 25DZ3008100, 2019SHZDZX01), the National Key R&D Program of China (grant no. 2022YFA1403202), the Quantum Science and Technology-National Science and Technology Major Project (grant no. 2024ZD0300103) and the Large Scientific Facility Open Subject of Songshan Lake Laboratory (grant no. DG23313511). Y.Z. was supported by the Youth Foundation of the National Natural Science Foundation of China (grant no. 12304173). Q.W. acknowledges support from the Research Grants Council of Hong Kong (ECS no. 24306223).

**Author contributions:** J.Z. conceived and supervised the project. L.C., E.Z., and Y.Z. synthesized the single-crystal samples. L.C., E.Z., B.C., F.L. and W.W. carried out sample characterizations. L.C., E.Z., Y.Z., and H.Z. performed the inelastic neutron scattering measurements with support from D.L.A. and T.J.W. Y. I. and L.C. performed elastic neutron scattering measurements. L.C., Y.H., W.W., Q.W., and J.Z. analyzed the data. J.Z. and L.C. wrote the manuscript. All authors provided comments for the manuscript.

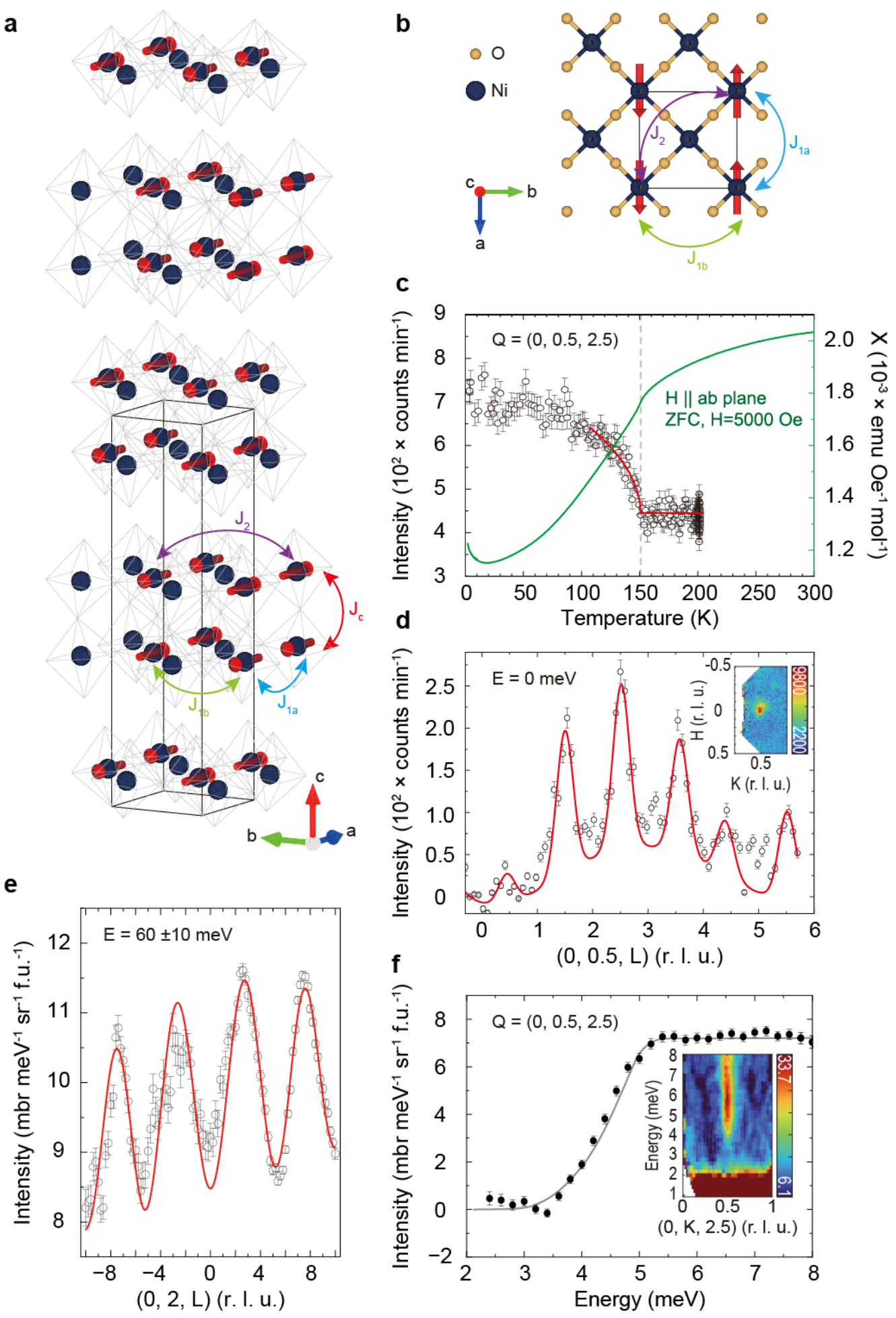

**Figure 1 | Magnetic structure, susceptibility, and spin dynamics of $La_3Ni_2O_7$. a,** Schematic illustration of the stripe-type magnetic structure. The moments on magnetic Ni ions and relevant exchange interactions are indicated in the figure. The unit cell is represented by a cuboid with dimensions of $a$ = 5.399 Å, $b$ = 5.440 Å, and $c$ = 20.631 Å. For clarity, only Ni ions are shown. Red arrows denote the ordered moments on magnetic Ni sites; non-magnetic Ni sites are drawn without arrows. **b,** Magnetic structure and exchange interactions within the $NiO_2$ plane. **c,** Temperature dependence of the zero-field-cooled (ZFC) magnetic susceptibility ($\chi$) measured under an external magnetic field of 5000 Oe applied parallel to the *ab* plane, together with the magnetic order parameter obtained from neutron diffraction. The intensity of the magnetic Bragg peak (0, 0.5, 2.5) follows an order-parameter-like behavior with a transition temperature of $T_N$ = 151 K and a critical exponent $\beta$ = 0.24±0.03. **d,** Magnetic Bragg peaks at $T$ = 3.6 K at $\boldsymbol{Q}$ = (0, 0.5, $L$), where the background measured at $T$ = 201 K has been subtracted. **e,** $L$-dependence of the magnetic excitation intensity at 60±10 meV, exhibiting a pronounced bilayer periodic modulation. Similar behavior was also observed at low energies (Supplementary Fig. S7). The intensity follows $I(q_x, q_y, q_z, \omega) = I_0(q_x, q_y, q_z, \omega)\sin^2(q_z d/2)$, where $d$ is the distance between adjacent $NiO_2$ layers. **f,** Energy dependence of low-energy spin excitations at $\boldsymbol{Q}$ = (0, 0.5, 2.5), revealing a spin gap of ~5 meV. The incoherent background away from the magnetic signal has been subtracted. The inelastic data were collected with ARCS and the elastic data were collected with TOPAN. Error bars in **c**–**f** represent one standard deviation (±1σ) propagated from neutron counting statistics. **Q** = ($H$, $K$, $L$) denotes the momentum-transfer vector; $q_x$, $q_y$ and $q_z$ are its components along the reciprocal-space directions; $\omega$ is the energy

transfer; and $I_0(q_x, q_y, q_z, \omega)$ is the intensity envelope excluding the bilayer-interference factor. mb, millibarns; f.u., formula unit; rlu, reciprocal lattice units.

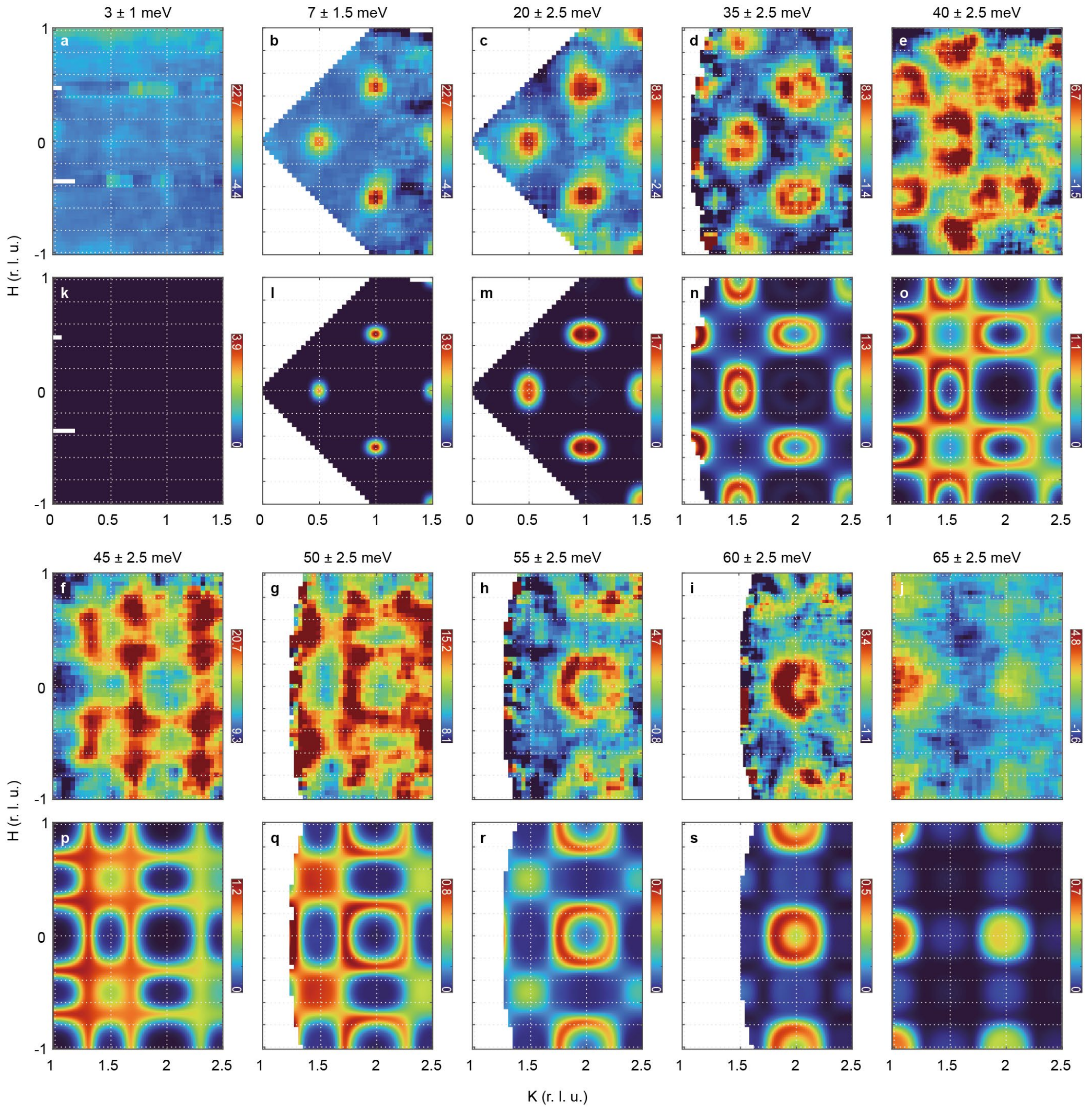

**Figure 2 | Momentum dependence of the spin excitations in $La_3Ni_2O_7$. a–j,** Constant-energy slices of the magnetic excitations measured at $T = 6$ K, integrated over energies of 3 ± 1 meV (**a**), 7 ± 1.5 meV (**b**), 20 ± 2.5 meV (**c**), 35 ± 2.5 meV (**d**), 40 ± 2.5 meV (**e**), 45 ± 2.5 meV (**f**), 50 ± 2.5 meV (**g**), 55 ± 2.5 meV (**h**), 60 ± 2.5 meV (**i**) and 65 ± 2.5 meV (**j**). The $|\boldsymbol{Q}|$-dependent isotropic background has been subtracted using an approach like that in ref. [53] for all panels except **f** and **g**, where the signals are broad and raw data are, therefore, shown. In **f** and **g**, a moderately enhanced background at large $|\boldsymbol{Q}|$ originates from the phonon scattering background. The measurements were performed on ARCS with incident neutron energies of 20 meV (**a**, **b**), 60 meV (**d**) or 120 meV (**c** and **e–j**). Symmetry-equivalent data were collected and averaged to improve statistical accuracy. **k–t,** Simulated constant-energy slices of the magnetic excitations, corresponding to the same reciprocal-space regions as the experimental data in **a–j**: 3 ± 1 meV (**k**), 7 ± 1.5 meV (**l**), 20 ± 2.5 meV (**m**), 35 ± 2.5 meV (**n**), 40 ± 2.5 meV (**o**), 45 ± 2.5 meV (**p**), 50 ± 2.5 meV (**q**), 55 ± 2.5 meV (**r**), 60 ± 2.5 meV (**s**) and 65 ± 2.5 meV (**t**). The color scale for the experimental data is in units of mb $meV^{-1}$ $sr^{-1}$ f.u.$^{-1}$, whereas that for the simulated slices is in arbitrary units.

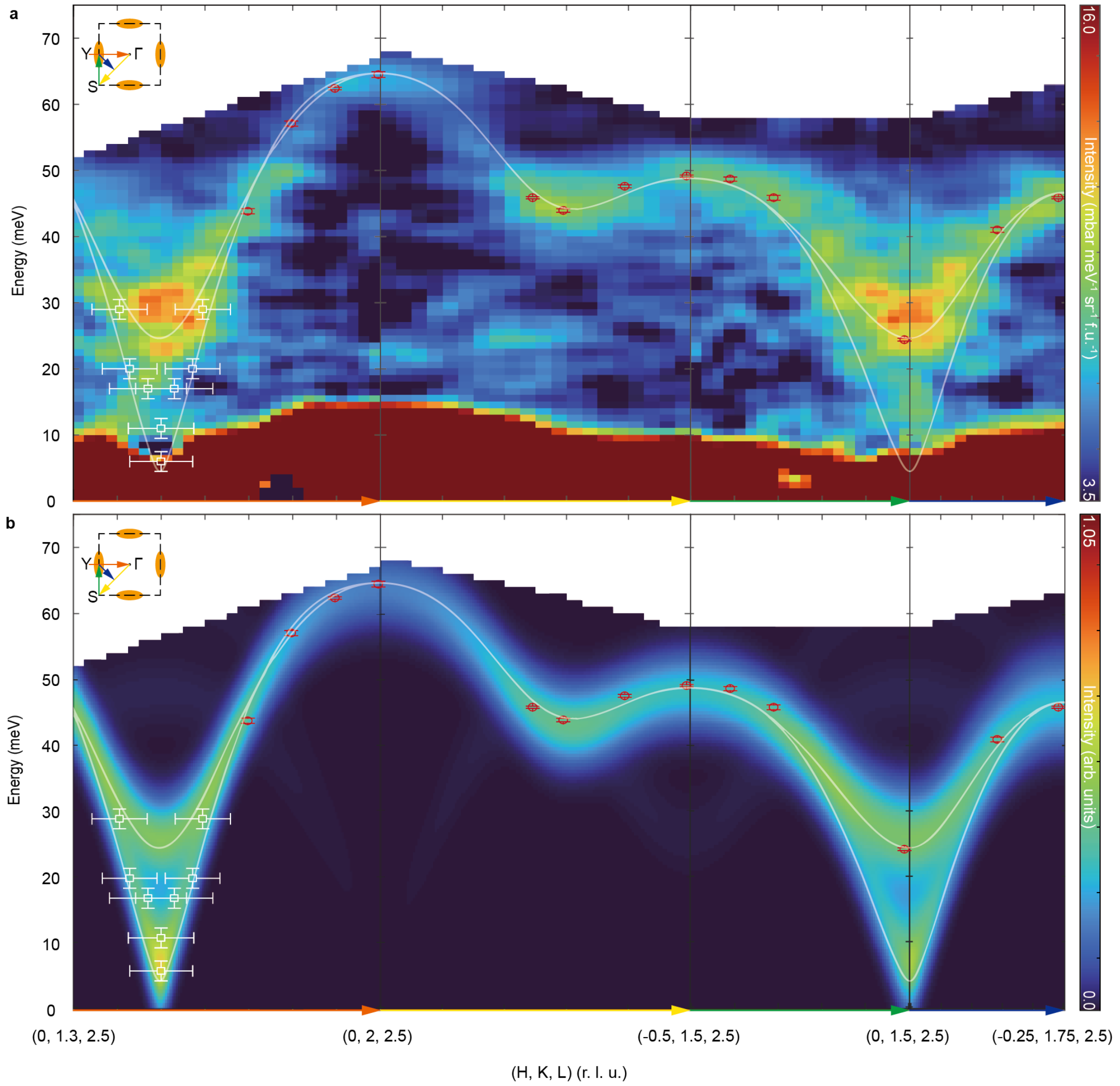

**Figure 3 | Measured and calculated spin-wave dispersions in $La_3Ni_2O_7$ at $T$ = 6 K. a,** Background-subtracted intensity map of the spin-excitation spectrum along the high-symmetry momentum directions (indicated by the arrows in the inset). The background was subtracted using the Shiver procedure. **b,** Simulated spin-excitation dispersion over the same region of reciprocal space as shown in **a**. The white solid line denotes the calculated dispersion relation. Red circles mark peak positions determined from fits for energy cuts (Supplementary Fig. S3). White squares indicate the peak positions extracted from Fig. 4 c–g. For the white squares, the vertical bars indicate the energy-integration ranges, and the horizontal bars represent the full-widths at half-maximum obtained from fits. For the red circles, the error bars indicate 1$\sigma$ uncertainties from the fit. The enhanced low-energy intensity below 10 meV in the measured spectra arises from incoherent scattering associated with the elastic line for incident energy $E_i$= 120 meV. Data were collected with ARCS.

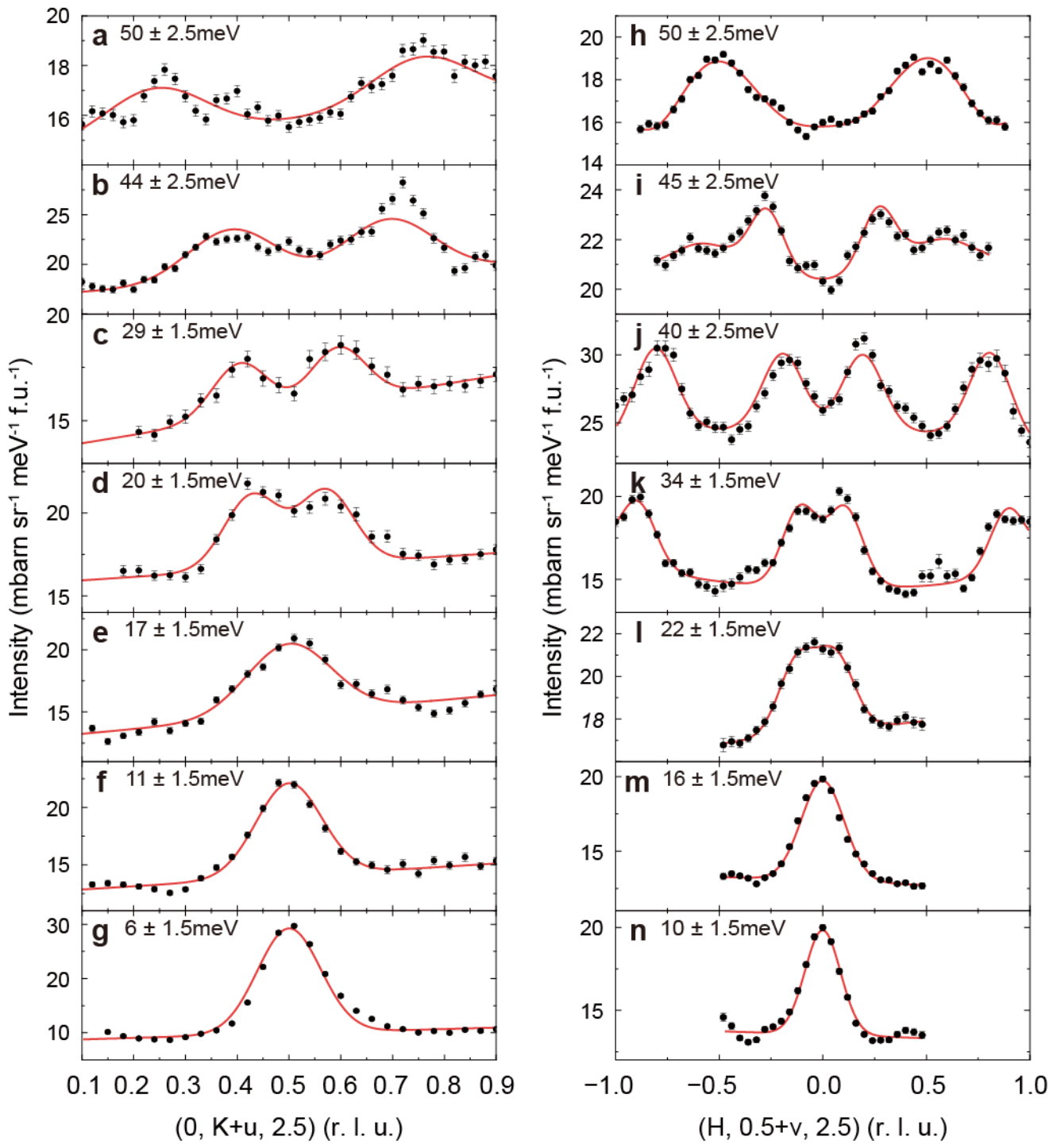

**Figure 4 | Constant-energy cuts of the spin excitations in $La_3Ni_2O_7$. a-g,** Constant-energy cuts of the spin excitation along the *K* direction at 6 K for the indicated energies: 50 ± 2.5 meV (**a**), 44 ± 2.5 meV (**b**), 29 ± 1.5 meV (**c**), 20 ± 1.5 meV (**d**), 17 ± 1.5 meV (**e**), 11 ± 1.5 meV (**f**) and 6 ± 1.5 meV (**g**). The parameter $u$ = 1 for **a** and **b** and $u$ = 0 for **c-g**. **h-n,** Constant-energy cuts along the *H* direction at 6 K: 50 ± 2.5 meV (**h**), 45 ± 2.5 meV (**i**), 40 ± 2.5 meV (**j**), 34 ± 1.5 meV (**k**), 22 ± 1.5 meV (**l**), 16 ± 1.5 meV (**m**) and 10 ± 1.5 meV (**n**).. The parameter $v$ = 1 for **h-k** and $v$ = 0 for **l-n**. Error bars in all panels represent one standard deviation ($\pm 1\sigma$) propagated from neutron counting statistics. Data were collected on ARCS.

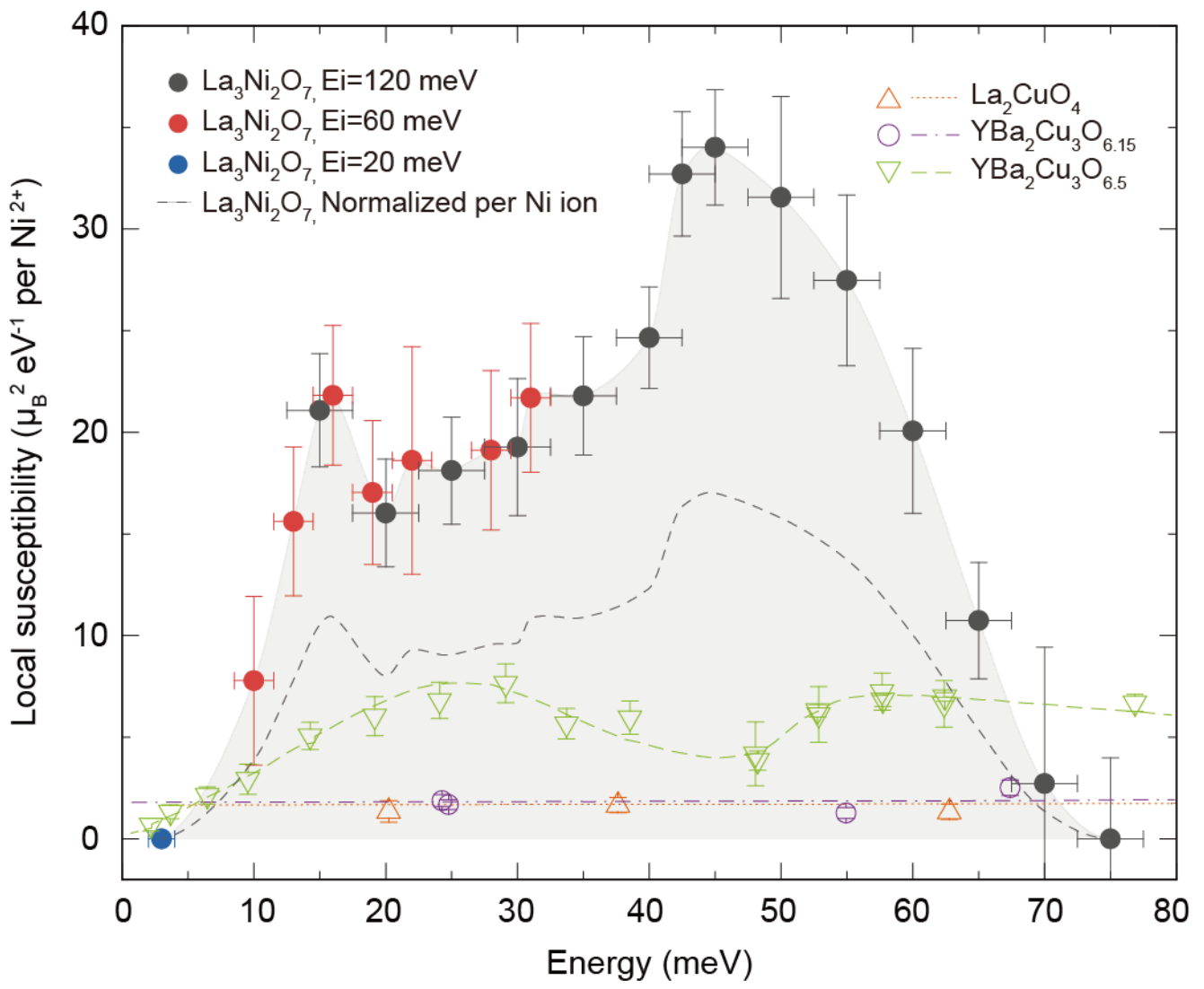

**Figure 5 | Energy dependence of the local dynamic susceptibility χ''(ω) in $La_3Ni_2O_7$.** Blue, red and black circles are from the measurements with incident energies $E_i$ = 20 meV, 60 meV and 120 meV at 6 K, respectively. The data were integrated over $1 < L < 6.2$ r.l.u., covering approximately one full bilayer-modulation period and including both acoustic and optical magnetic signals. The low- and mid-energy spectral weight was measured with several incident energies, and only data points that were mutually consistent between different $E_i$ settings and free from identifiable phonon or sample-environment contamination were used to construct $\chi''(\omega)$. The horizontal bars represent the energy integration ranges, and the vertical bars denote the statistical uncertainties, representing one standard deviation (±1σ) propagated from neutron counting statistics. The shaded area shows $\chi''(\omega)$ normalized per magnetic $Ni^{2+}$ site in the stripe model. The dashed gray curve shows the same spectrum rescaled per total planar Ni site; in the nominal mixed-valence stripe picture, half of the Ni sites carry sizable magnetic moments, so this curve is smaller by a factor of two. The inelastic neutron scattering data were collected with ARCS. The cuprate data points were extracted from the literature and normalized per Cu ion: open triangles and circles denote cuprate parent compounds measured at 295 K (refs. [54,55]), while open inverted triangles denote the normal state of the $YBa_2Cu_3O_{6.5}$ superconductor at 60 K (ref. [56]). For $YBa_2Cu_3O_{6.15}$ and $YBa_2Cu_3O_{6.5}$, the acoustic and optical branches are summed.

## Methods

### Growth and characterization of $La_3Ni_2O_7$ single crystals

The polycrystalline precursor of $La_3Ni_2O_7$ was synthesized via a conventional solid-state reaction. High-purity $La_2O_3$ and NiO powders (Aladdin, 99.99%) were weighed in stoichiometric proportions, thoroughly ground, and subsequently calcined at 1400 K for 24 hours. We conducted another cycle of grinding and calcination to ensure homogeneity. The resulting powder was then hydrostatically pressed into a cylindrical rod (approximately 6 mm in diameter and 10 cm in length) under a pressure of 300 MPa. The rod was sintered at 1400 K for 12 hours in air so that it attained sufficient density and mechanical strength for the crystal growth process. Single-crystal growth was carried out in a vertical optical floating-zone furnace (Model HKZ, SciDre) under an oxygen pressure of 14-16 bar using a 5-kW xenon arc lamp as the heat source. The feed rod was translated through the molten zone at a rate of 5 mm $h^{-1}$.

Phase purity and crystal structure were verified by X-ray diffraction (Cu–Kα radiation, D8 Discover diffractometer, Bruker; Supplementary Fig. S1). Magnetic and transport properties were characterized using Quantum Design MPMS-3 SQUID magnetometer and Physical Property Measurement System. Zero-field-cooled magnetization measurements between 2 and 300 K revealed antiferromagnetic ordering below ~151 K. The resistivity exhibited metallic behavior with a clear anomaly near the magnetic transition temperature (Supplementary Fig. S1). Thermogravimetric analysis was performed on NETZSCH STA 449F3 and STA 449F1 instruments (Supplementary Fig. S11). Chemical compositions were analysed by an energy-dispersive X-ray spectrometer attached to a field-emission scanning electron microscope (Sigma 300, Zeiss) (Supplementary Fig. 10).

**Neutron scattering experiments**

Inelastic neutron scattering measurements were performed with the ARCS spectrometer at Oak Ridge National Laboratory and the MAPS spectrometer at ISIS Neutron and Muon Facility. Twenty-five $La_3Ni_2O_7$ single crystals, with a total mass of ~1.26 g, were co-aligned in the (*0*, *K*, *L*) scattering plane. Experiments on ARCS were conducted in a closed-cycle refrigerator down to 6 K using incident energies $E_i$ = 20 meV, 60 meV, 120 meV, and 200 meV. Experiments on MAPS were performed in a closed-cycle refrigerator from 6 K to 300 K with $E_i$ = 40 meV, 60 meV, 100 meV, 120 meV, 180 meV, and 200 meV. The magnetic scattering intensities were placed on an absolute scale by normalization to acoustic phonons, enabling absolute calibration of the imaginary part of the dynamic susceptibility, $\chi''(\omega)$ and allowing quantitative comparison with that of cuprates and other superconductors. Data reduction and analysis were performed using the HORACE and Shiver software packages, which we used to slice and project the four-dimensional dataset into (*H, K, L, E*) space.

The elastic neutron scattering experiment was performed using the triple-axis spectrometer TOPAN installed at the 6G horizontal beam port of the Japan Research Reactor 3 (JRR-3). We used neutrons with an energy of 13.4 meV. A vertically focusing pyrolytic graphite monochromator and analyzer were employed. The collimation was set to blank-60′-60′-60′. To suppress higher-order contamination, pyrolytic graphite filters were placed in front of both the monochromator and analyzer. The sample was mounted on an aluminum plate using CYTOP and sealed in an aluminum can together with helium exchange gas. The sample was cooled using a water-cooled Gifford-McMahon refrigerator, and measurements were carried out in a temperature range of approximately 3.6 K to 200 K.

# Supplementary Information

## Table of Contents

## Supplementary Text

## Supplementary Figures

## Supplementary References

**Sample characterization**

To further characterize the crystals used in the neutron-scattering measurements, we performed complementary chemical and thermogravimetric analyses on representative crystals from the co-aligned array. Energy-dispersive X-ray spectroscopy (EDX/EDS) was carried out at multiple positions to evaluate the cation homogeneity (Supplementary Fig. S10). The averaged La:Ni ratio was found to be close to 1.5, consistent with the nominal $La_3Ni_2O_7$ composition within experimental uncertainty. No significant spatial variation of the La:Ni ratio was observed within the probed regions. Thermogravimetric analysis (TGA) was used to estimate the oxygen content (Supplementary Fig. S11). From the mass change measured during reduction under an $H_2$/Ar atmosphere, the oxygen stoichiometry is estimated to be approximately $La_3Ni_2O_{6.94}$, indicating that the crystals are close to the nominal oxygen content, with only a minor oxygen deficiency. For the co-aligned 25-crystal array used in the neutron-scattering measurements, the mosaic spread was estimated from transverse scans through nuclear Bragg peaks. The transverse scan through the (0,4,0) nuclear Bragg peak gives an in-plane azimuthal mosaic width with a full width at half maximum (FWHM) of about 1.98°, whereas the transverse scan through the (0,0,12) peak gives a *c*-axis tilt distribution of about 5.30° (Supplementary Fig. S13). The broader out-of-plane width reflects the greater sensitivity of the plate-like crystals to small *c*-axis tilts during co-alignment, whereas the in-plane azimuthal alignment is more tightly constrained. These values show that the co-aligned array is sufficiently well aligned for the momentum-resolved neutron measurements.

More importantly, the inelastic magnetic spectra provide an internal consistency check supporting the assignment of the measured magnetic signal to the bilayer $La_3Ni_2O_7$ phase. The most phase-sensitive feature is the observed $L$-dependent intensity modulation, which follows the bilayer periodicity of $La_3Ni_2O_7$. This bilayer-specific modulation is distinct from the out-of-plane periodicities associated with $La_4Ni_3O_{10}$, which has a trilayer stacking sequence, and the 1313 polymorph, which contains alternating monolayer and trilayer blocks. The combination of bulk-sensitive neutron characterization, the $La_3Ni_2O_7$-specific bilayer modulation, the in-plane magnetic wave vector, finite spin gap and dispersion, and complementary chemical analyses supports the assignment of the measured magnetic signal to the bilayer $La_3Ni_2O_7$ phase.

**Absolute unit calibration**

The absolute-unit calibration was performed following the standard procedure[1]. The measured neutron intensity was converted to the imaginary part of the dynamic susceptibility $\chi''(\boldsymbol{Q},\omega)$ using the standard magnetic neutron scattering cross-section with the isotropic $Ni^{2+}$ magnetic form factor included. To simplify the analysis, we adopted an isotropic approximation $\chi''(\boldsymbol{Q},\omega) = Tr\left(\chi''_{\alpha\beta}(\boldsymbol{Q},\omega)\right)/3$, a treatment widely employed for magnetically ordered cuprate[2,3] and iron-based parent compounds[4]. The magnetic spectra were placed on an absolute intensity scale by normalization to an acoustic phonon measured near the nuclear Bragg position Q = (0, 0, 10) (Supplementary Fig. S12). The longitudinal phonon intensity can be described by the one-phonon neutron cross section:

$$\frac{d^2\sigma}{d\Omega dE} = \frac{1}{d\omega/dq}\cdot N\cdot\frac{\hbar^2}{E(\boldsymbol{q})}\left(\frac{k_f}{k_i}\right)\frac{1}{2}n(q)\frac{|\boldsymbol{G}|^2}{M}e^{-2W}\cdot|F_N(\boldsymbol{G})|^2$$

Where $\frac{d\omega}{dq}$ is the phonon dispersion velocity at the scan position, $N$ is the number of crystallographic unit cells in the sample, and $M$ is the mass of one crystallographic unit cell used in the phonon structure-factor calculation. The factor $n(q)$ is the phonon population factor for neutron energy-loss scattering. $\boldsymbol{G}$ is the reciprocal lattice vector of the Bragg peak corresponding to the measured phonon. $F_N(G)$ is the acoustic phonon structure factor. The measured acoustic phonon intensities are compared with the calculated one-phonon cross sections to place the neutron scattering intensity on an absolute scale. The absolute normalization is then applied to the magnetic scattering, after background subtraction and correction for the $Ni^{2+}$ magnetic form factor, to obtain local dynamic susceptibility $\chi''(\omega)$. This procedure provides an internally consistent absolute normalization for comparing the magnetic spectral weight of $La_3Ni_2O_7$ with cuprates and other correlated magnetic materials.

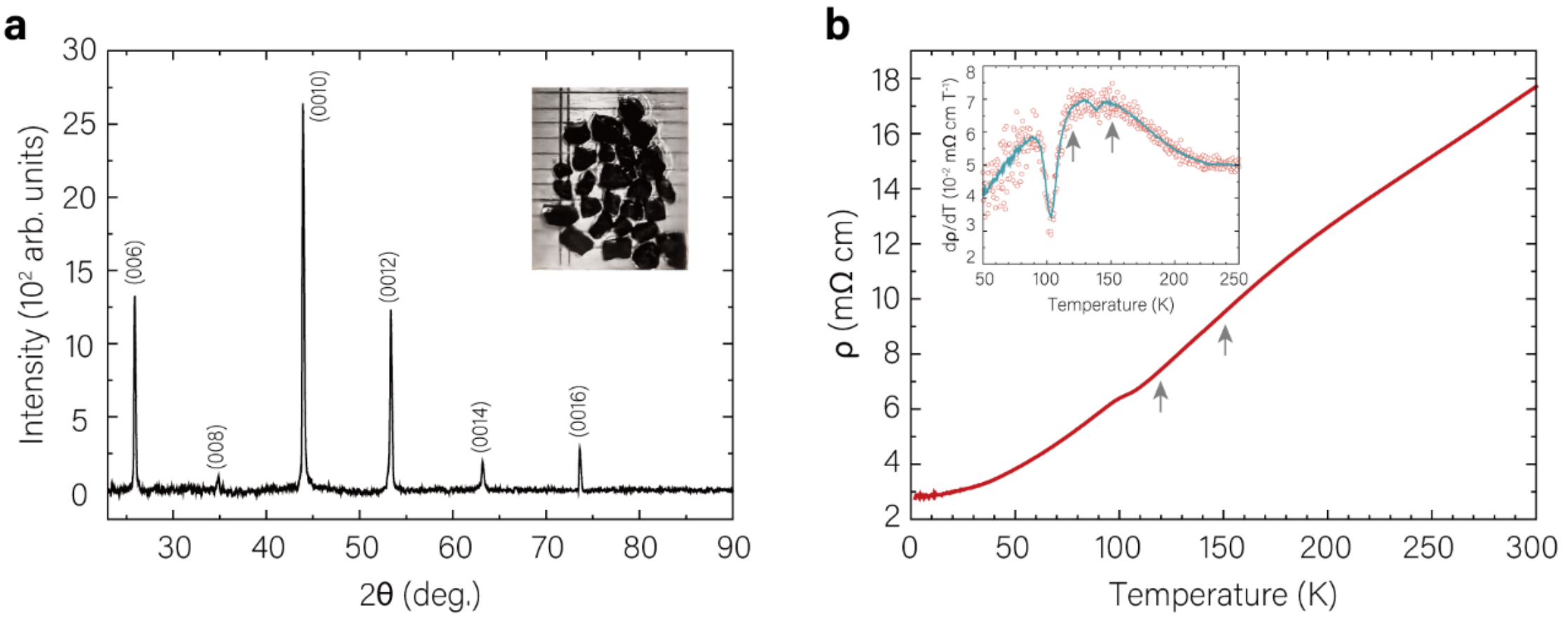

**Fig. S1 | X-ray diffraction (XRD) and resistivity measurements of $La_3Ni_2O_7$ single crystals. a,** XRD pattern collected from the *ab*-plane of a $La_3Ni_2O_7$ single crystal, showing sharp reflections and the absence of detectable secondary phases. Data were obtained using a Bruker D8 Discover diffractometer with Cu $K_\alpha$ radiation. The inset displays a photograph of coaligned $La_3Ni_2O_7$ single crystals for inelastic neutron scattering measurements. **b,** Temperature dependence of the *ab*-plane resistivity in $La_3Ni_2O_7$. The derivative (inset) exhibits two anomalies at $T_{CDW} \approx 120$ K and $T_N \approx 151$ K, corresponding to the onset of putative charge-density-wave and magnetic phase transitions, respectively.

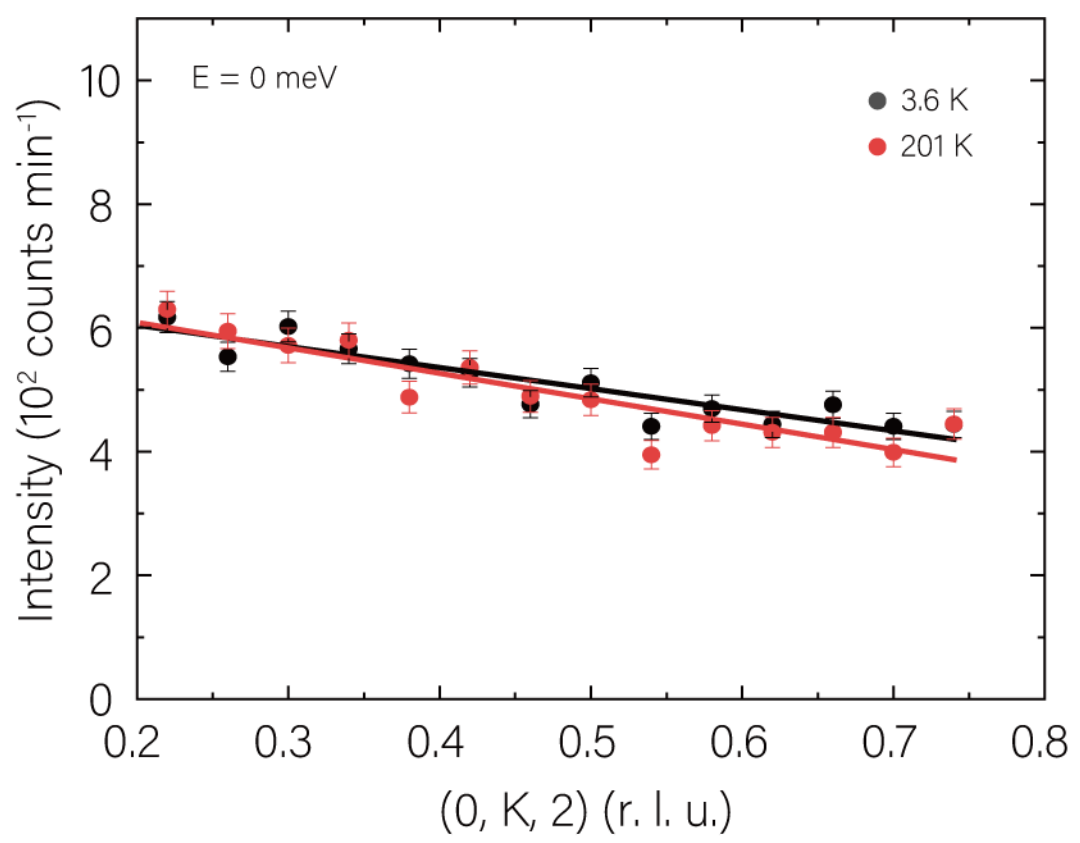

**Fig. S2 | Elastic neutron diffraction near (0, 0.5, 2) below and above $T_N$ in $La_3Ni_2O_7$.** Elastic neutron scattering measurements on TOPAN reveal no magnetic Bragg peak at (0, 0.5, 2) at 3.6 K, which corresponds to the magnetic propagation vector $\boldsymbol{q}$ = (0, 0.5, 0).

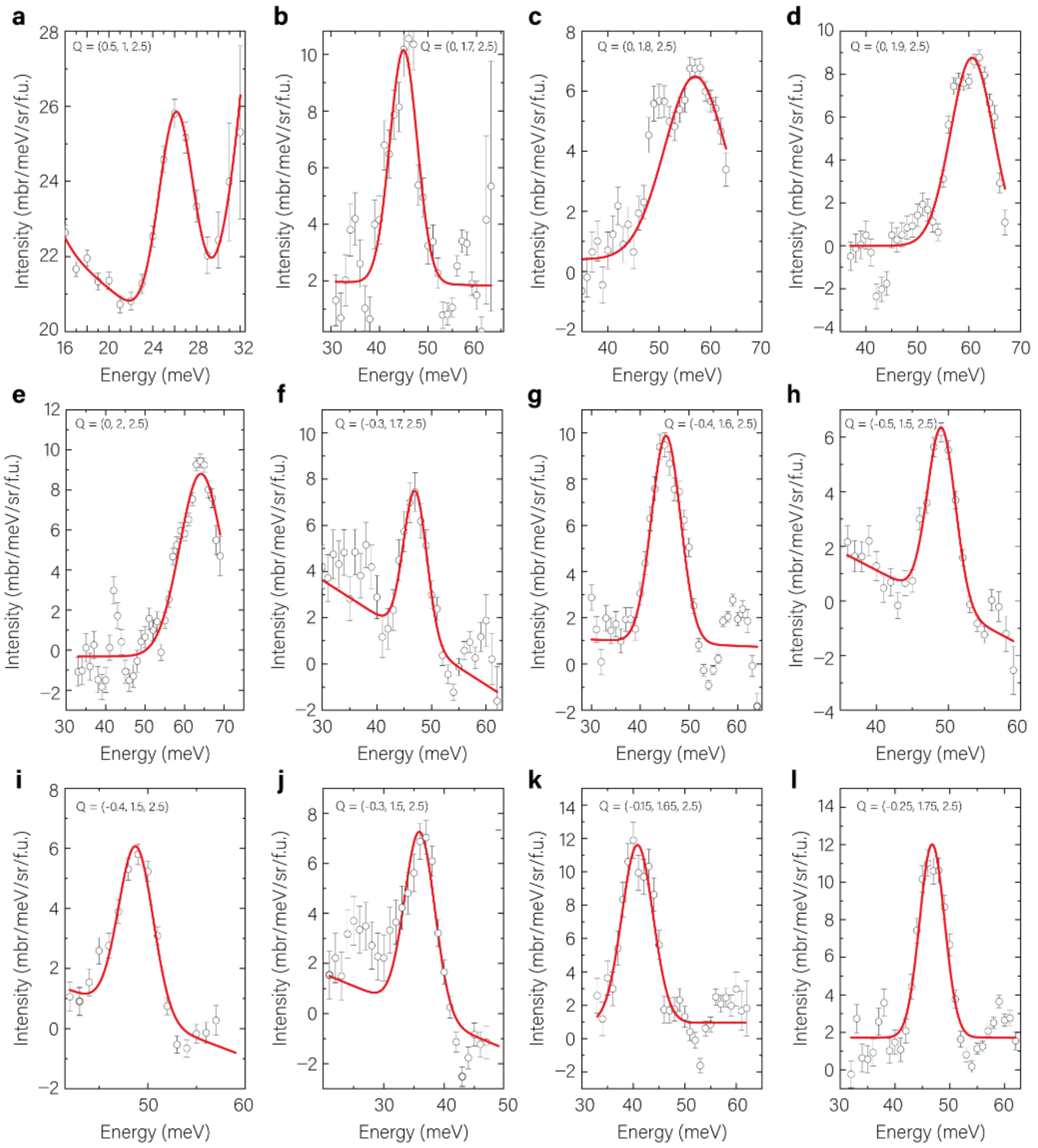

**Fig. S3 | Energy cuts at different positions in reciprocal space, as indicated.** A common energy-dependent background has been subtracted. Gaussian fitting is used to quantitatively determine the dispersion energy. The data were collected at 6 K on the ARCS spectrometer.

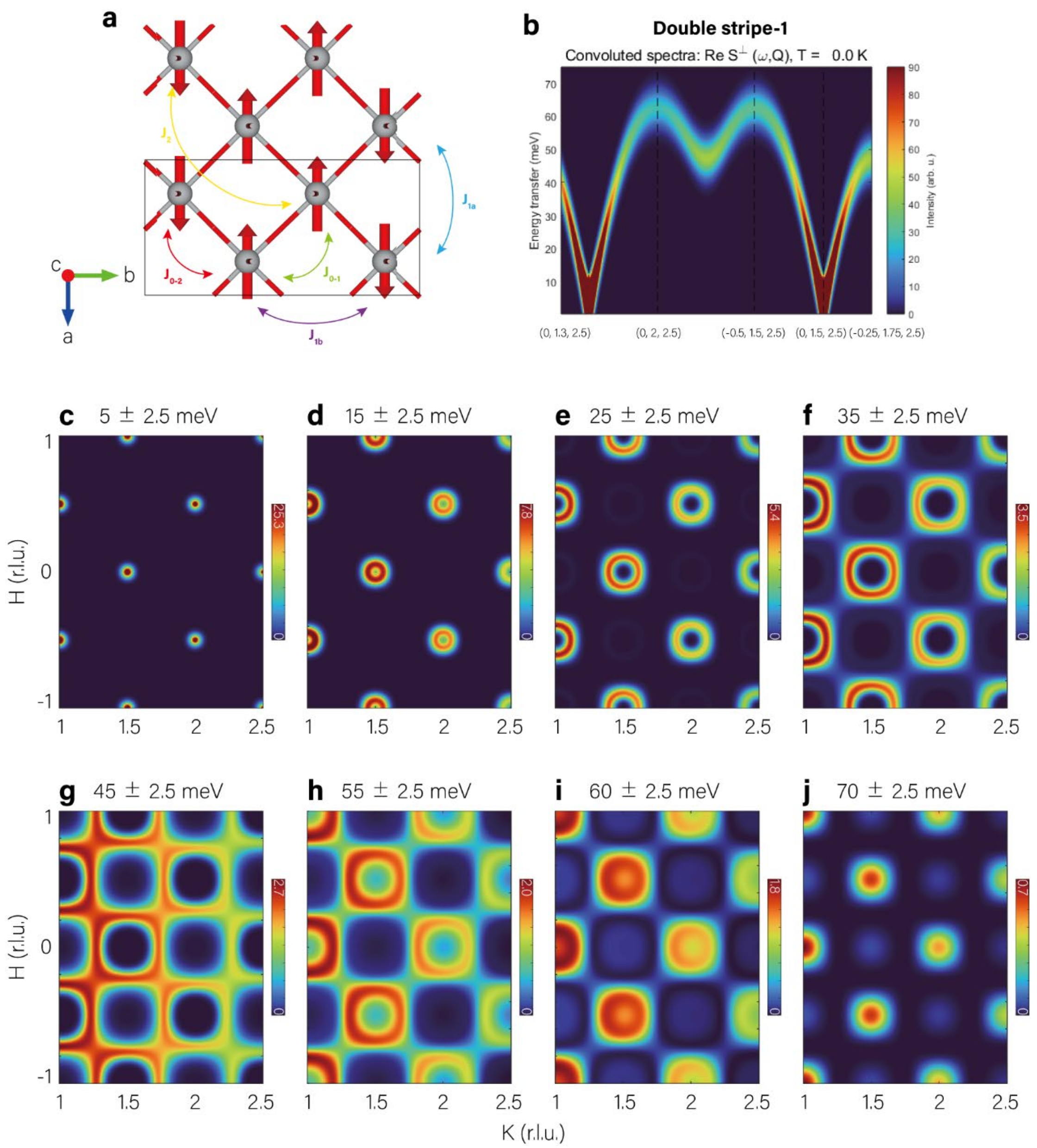

**Fig. S4 | Isotropic double stripe model of $La_3Ni_2O_7$ (Double Stripe 1)**. **a,** Schematic illustration of the in-plane exchange interactions in the double-stripe magnetic structure. **b,** Simulated energy–momentum (*E*-***Q***) spectrum for the double-stripe model with exchange parameters $SJ_c = 60.89$, $SJ_{0-1} = 0$, $SJ_{0-2} = 0$, $SJ_{1a} = 0$, $SJ_{1b} = 0$, $SJ_2 = 4$ meV. No softening of the magnetic scattering near (-0.5, 1.5, 2.5) is observed. **c-j,** Simulated in-plane slices at the indicated energies showing an isotropic scattering pattern in the *ab* plane, in clear contrast to the anisotropic features observed experimentally. The color-bar is in arbitrary units.

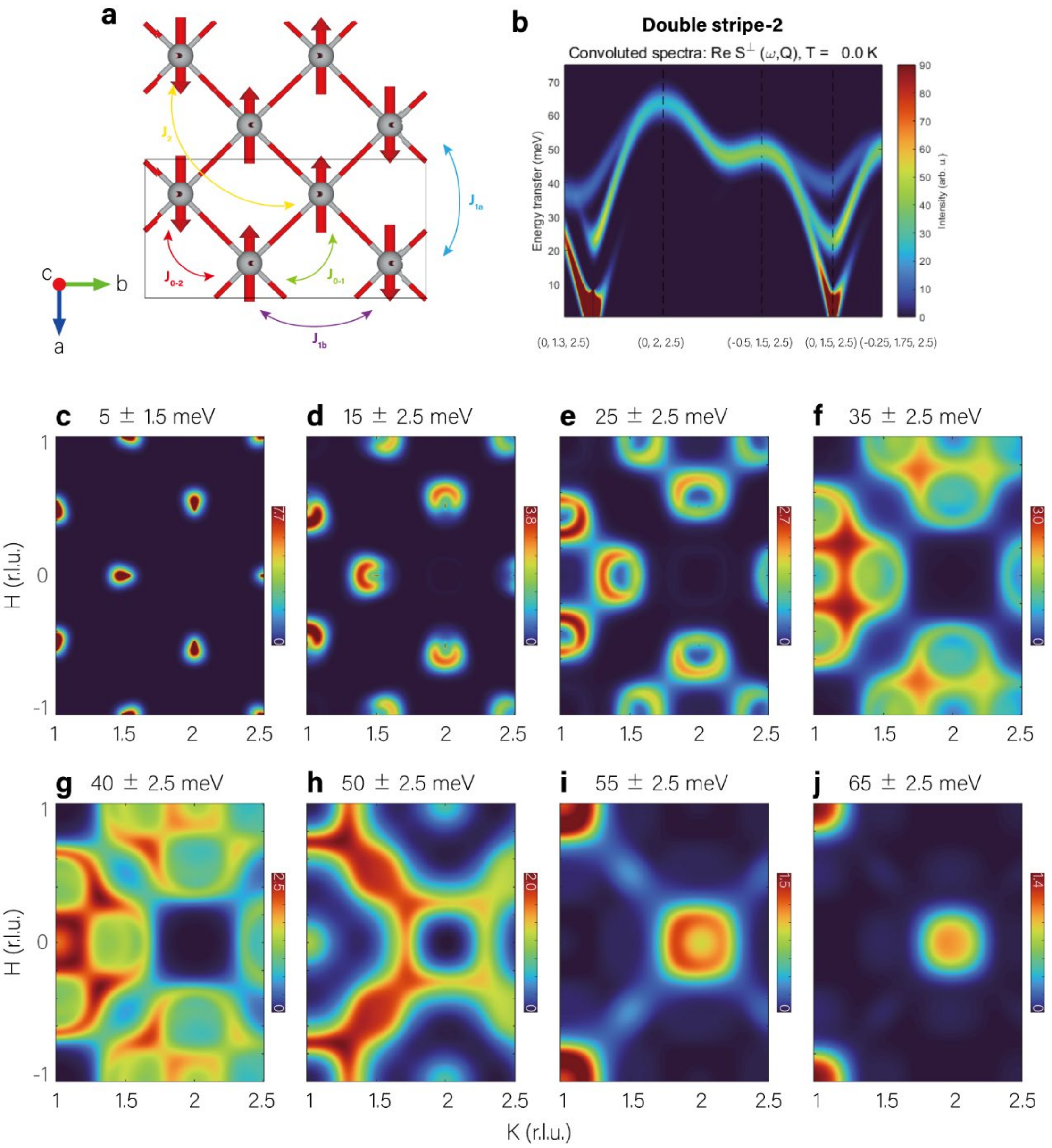

**Fig. S5 | Anisotropic double stripe model in $La_3Ni_2O_7$ (Double stripe-2)**. **a,** Schematic illustration of the in-plane exchange interactions in the double-stripe magnetic structure. **b,** Simulated energy–momentum (*E*-**Q**) spectrum for the double-stripe model with exchange parameters $SJ_c = 55.89$, $SJ_{0\text{-}1} = 2$, $SJ_{0\text{-}2} = 3$, $SJ_{1a} = -0.4$, $SJ_{1b} = 2$, $SJ_2 = 2$ meV. **c-j,** Simulated in-plane slices at indicated energies. The color-bar is in arbitrary units.

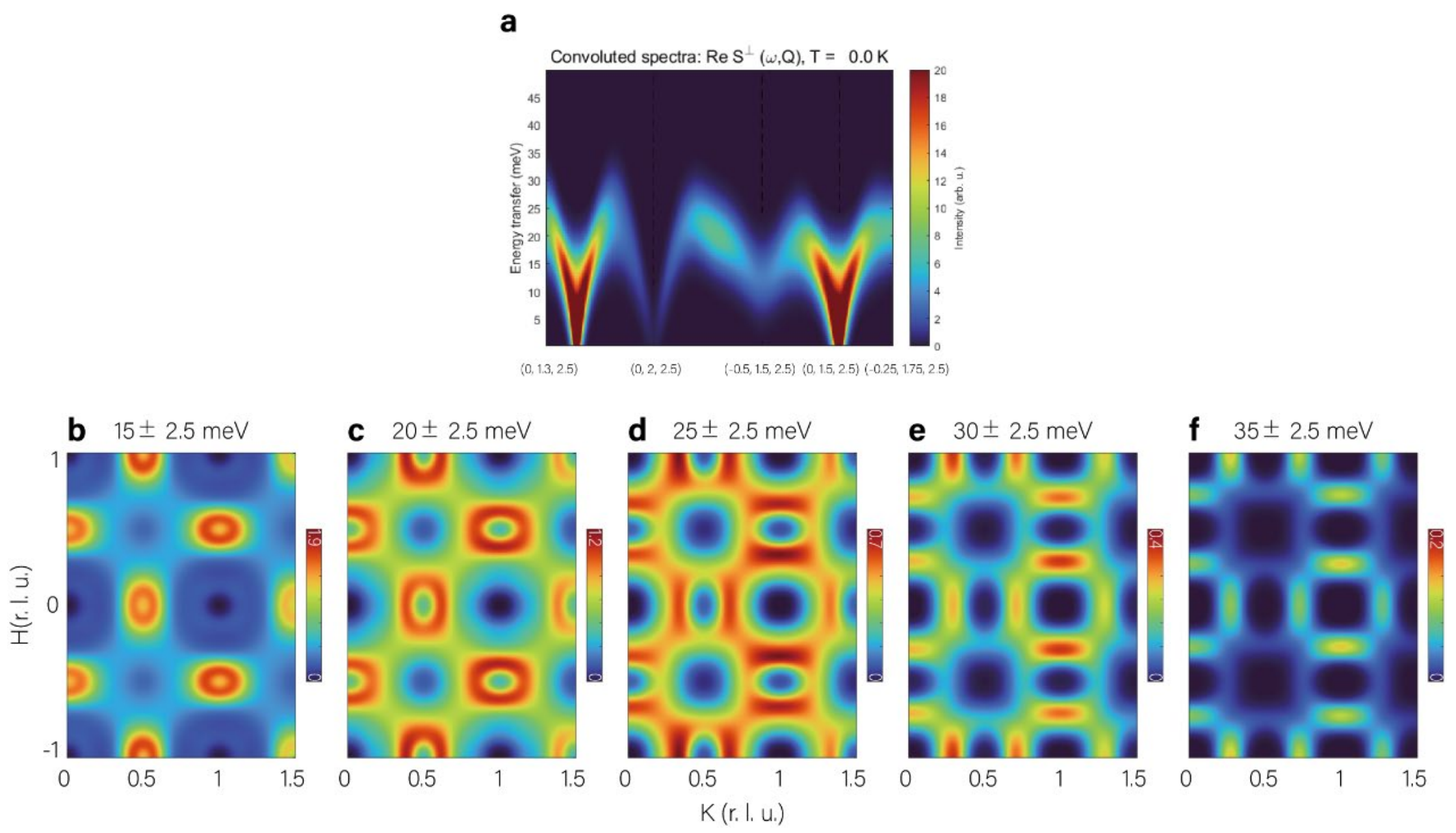

**Fig. S6 | Simulated magnetic excitation spectra of the single-stripe magnetic order with only in-plane exchange interactions. a,** Simulated energy–momentum (*E*-***Q***) spectrum for the single stripe model with exchange parameters $SJ_c = 0$, $SJ_2 = 4.63$, $SJ_{1b} = 3.71$, $SJ_{1a} = 2.36$ meV. **b-f,** Simulated in-plane slices at indicated energies. The color-bar is in arbitrary units.

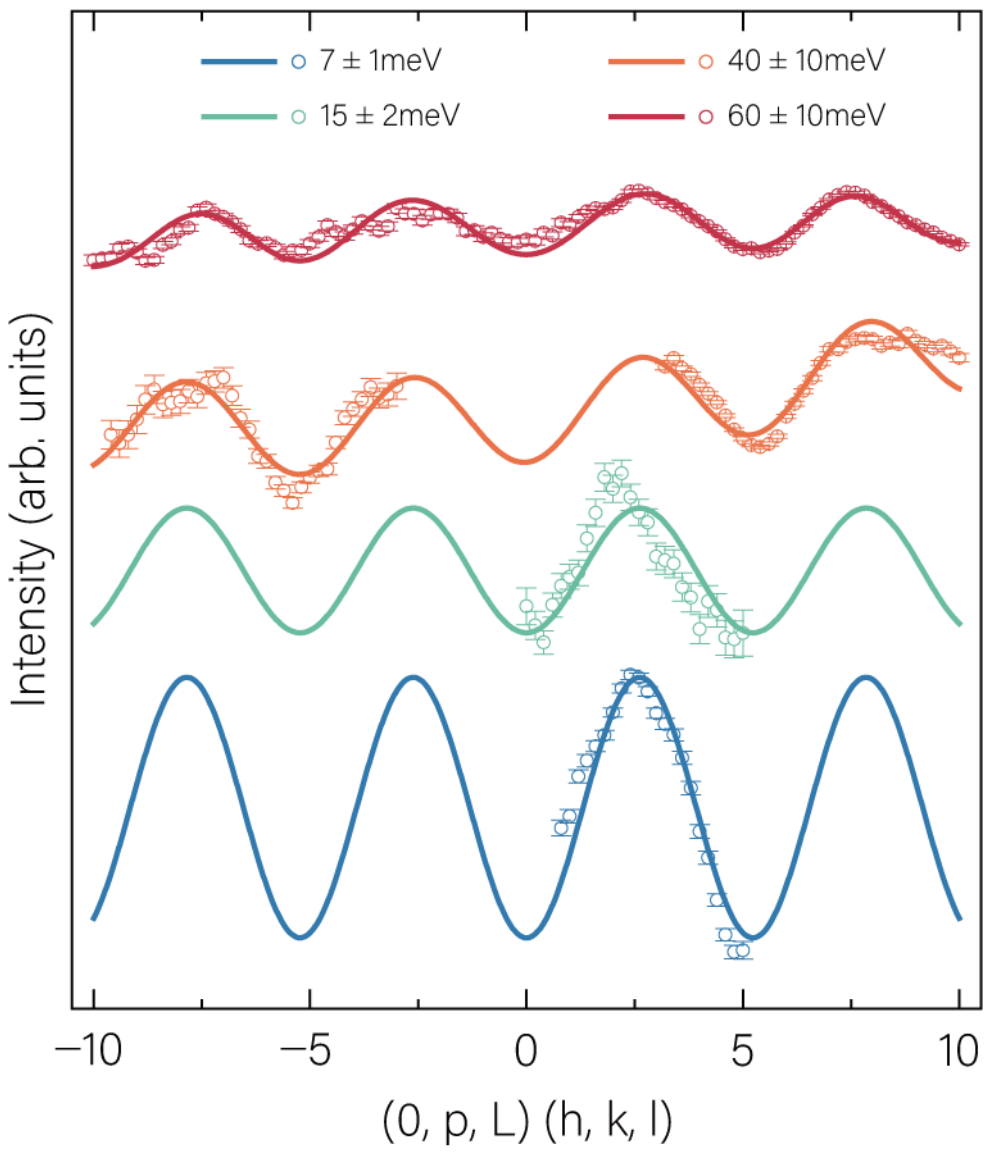

**Fig. S7 | *L* modulation of the magnetic excitation intensity for the acoustic branch at various energies.** The intensity follows $I(q_x, q_y, q_z, \omega) = I_0(q_x, q_y, q_z, \omega)\sin^2(q_z d/2)$, where $d$ is the spacing between adjacent $NiO_2$ layers. Here, p = 0.5, 0.5, 1.5 and 2 for E = 7, 15, 40 and 60 meV respectively. The 7, 40 and 60 meV data were measured on ARCS and the 15 meV data were measured on MAPS. Data were vertically offset for clarity.

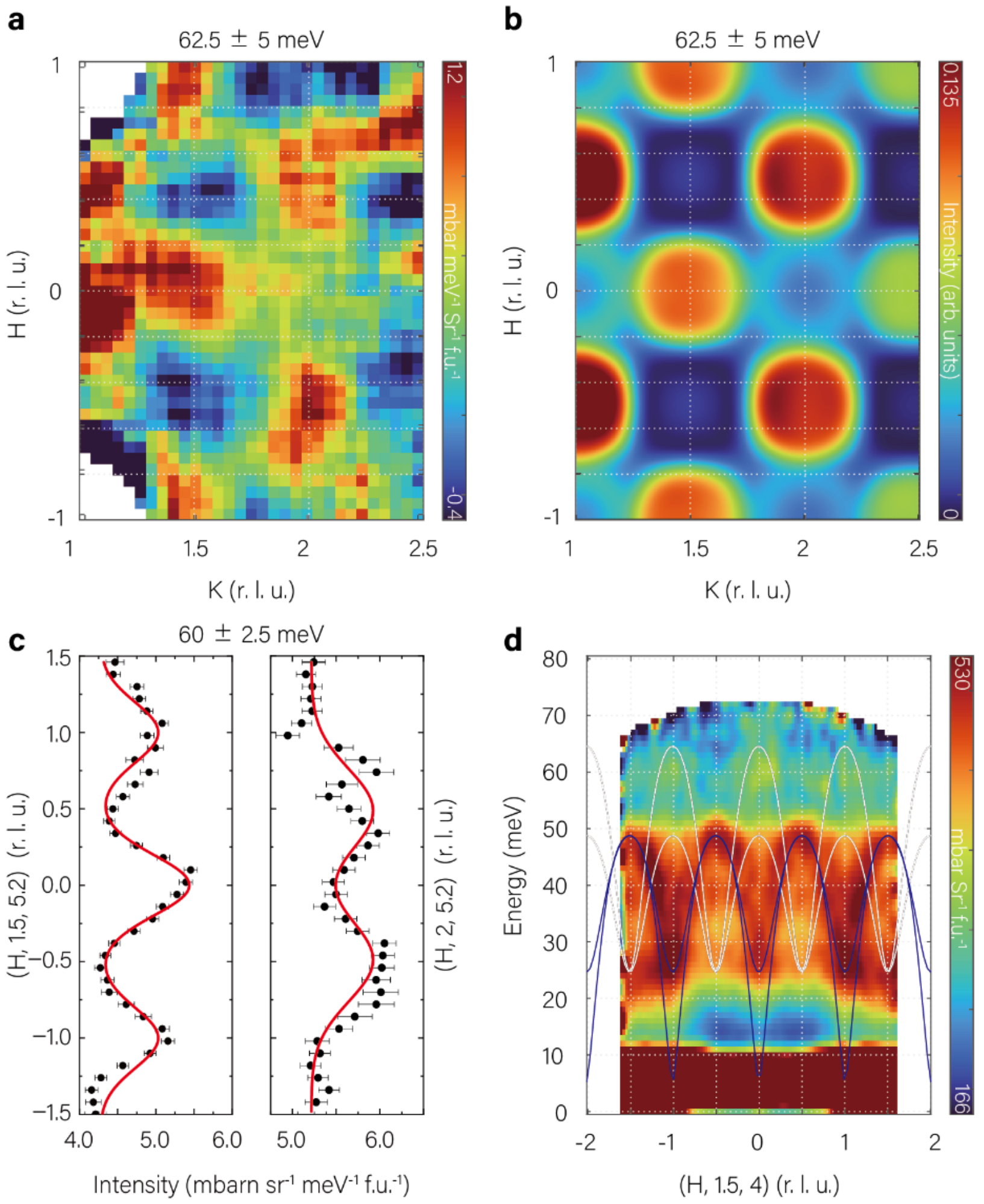

**Fig. S8 | Optical magnetic excitation branch of $La_3Ni_2O_7$. a,** Constant-energy slice of the optical magnetic excitation branch, integrated around $L$=5.2, where the optical mode shows its strongest intensity modulation. **b,** Simulated constant-energy slice of the optical magnetic excitation branch over the same reciprocal-space region as in **a**, calculated using the spin-wave model described in the main text. **c,** Constant-energy cuts of optical magnetic excitation along the $H$ direction at 6 K. **d,** Energy-momentum ($E$-$\boldsymbol{Q}$) dispersion along the $H$ direction. The scattering intensity is multiplied by the energy transfer $E$ to facilitate visualization, and the data are integrated over $L \in [1, 7]$ (r.l.u.). Both acoustic and optical branches are visible. Deep-blue lines indicate the acoustic branch, and white lines indicate the optical branch.

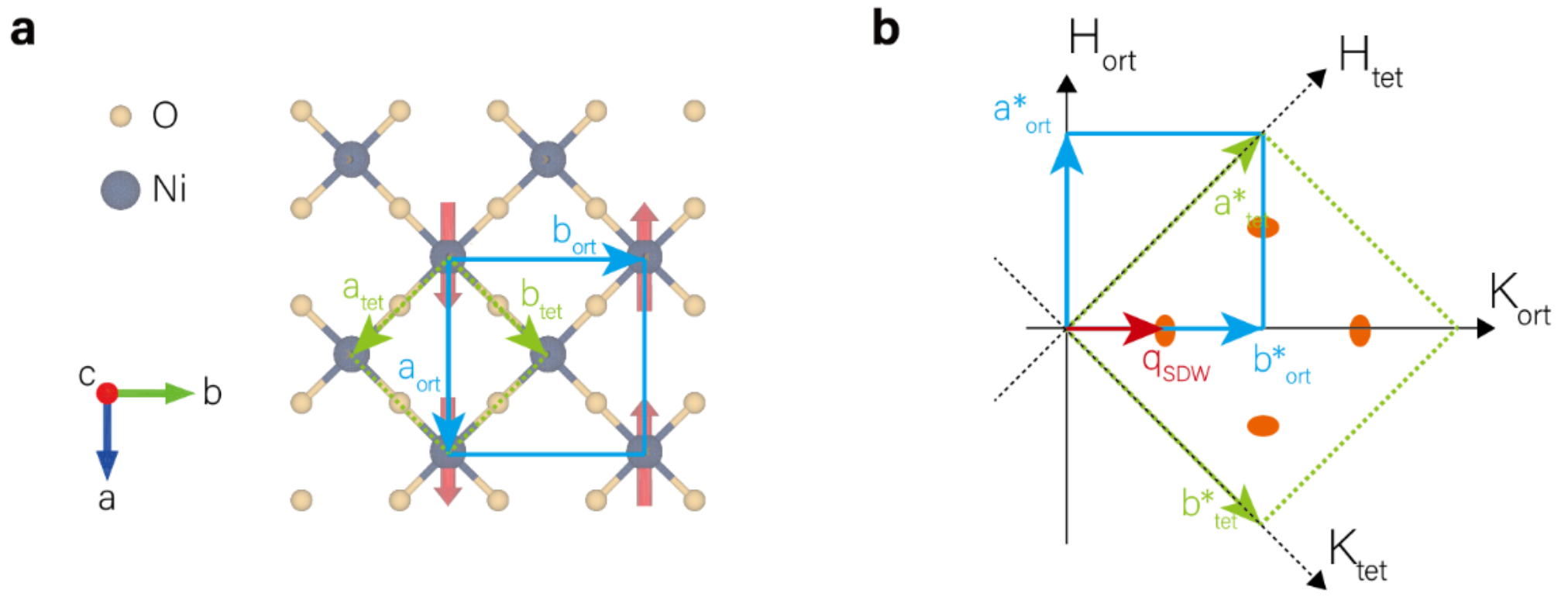

**Fig. S9 | a,** In-plane unit cells in the orthorhombic and tetragonal representations. The orthorhombic representation has lattice parameters $a$ = 5.399 Å, $b$ = 5.440 Å, and $c$ = 20.631 Å, whereas the tetragonal representation has $a = b \approx 3.833$ Å and $c$ = 20.631 Å. **b,** Reciprocal lattices corresponding to the two representations. In the orthorhombic setting, the magnetic ordering vector is $\boldsymbol{q}$ = (0, 0.5, 0.5), whereas in the tetragonal setting it is $\boldsymbol{q}$ = (0.25, 0.25, 0.5). The orthorhombic representation is used throughout this work.

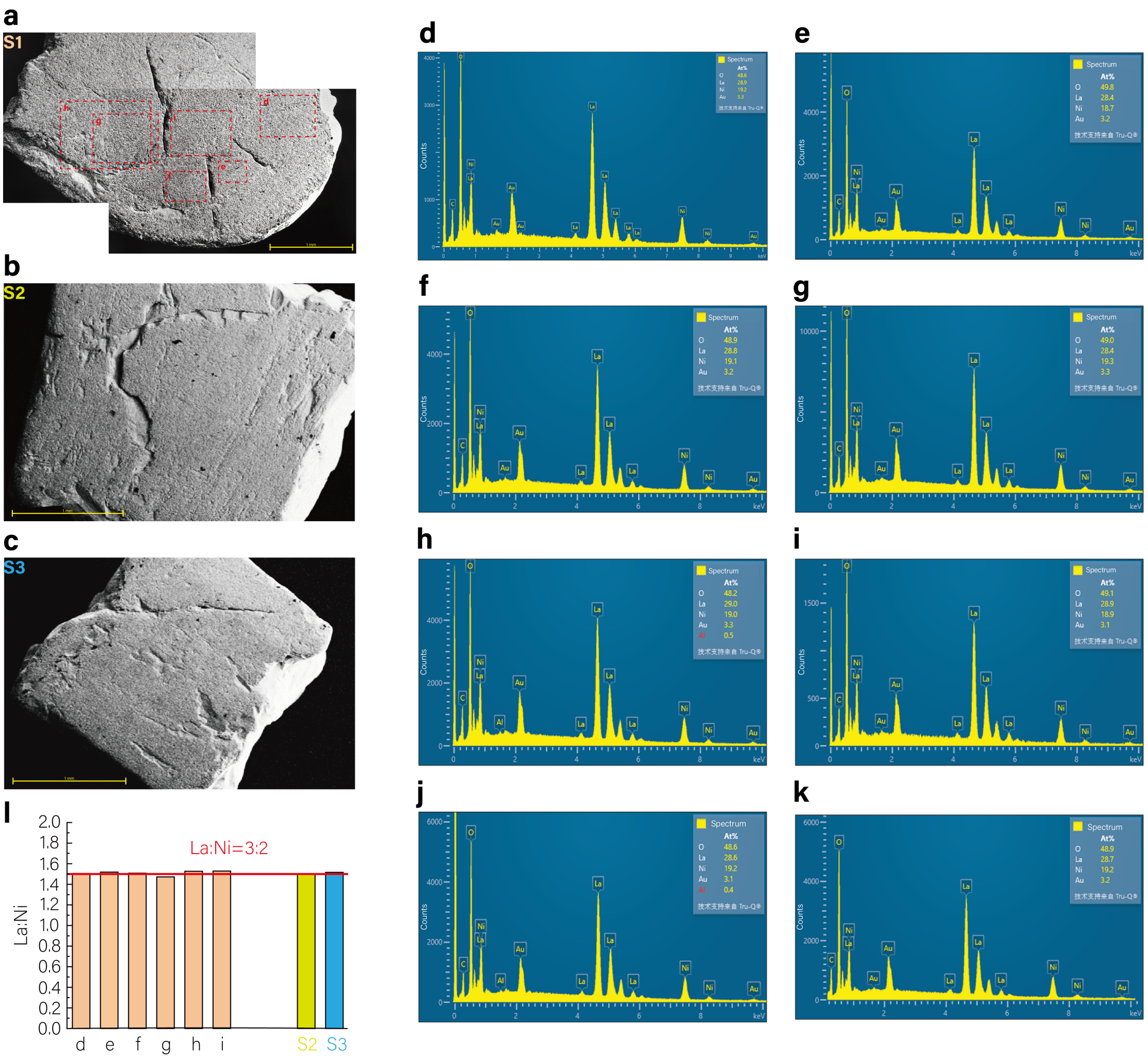

**Fig. S10 | Energy-dispersive X-ray spectroscopy characterization of $La_3Ni_2O_7$ single crystals.** **a–c**, Scanning electron microscopy (SEM) images of representative $La_3Ni_2O_7$ crystals S1–S3 used in the neutron scattering measurements. The exposed surfaces correspond to the *ab* plane, which was cut with a diamond saw during sample preparation and co-alignment. **d–i,** EDX spectra collected from six randomly selected regions on sample S1, marked by the red dashed boxes in **a**. **j,k,** EDX spectra collected from samples S2 and S3, respectively. The C signal mainly arises from the conductive tape, the Au signal comes from the thin surface coating used to reduce charging, and the weak Al peaks in **h** and **j** are attributed to the aluminum sample holder. Oxygen was not used for quantitative stoichiometry in the EDX analysis because of possible surface and substrate contributions. **l**, La/Ni atomic ratios extracted from the six measured regions of sample S1, together with the averaged La/Ni ratios of samples S2 and S3. The La/Ni ratios show only minor spatial and sample-to-sample variation and remain close to the ideal stoichiometric value of 1.5 for $La_3Ni_2O_7$.

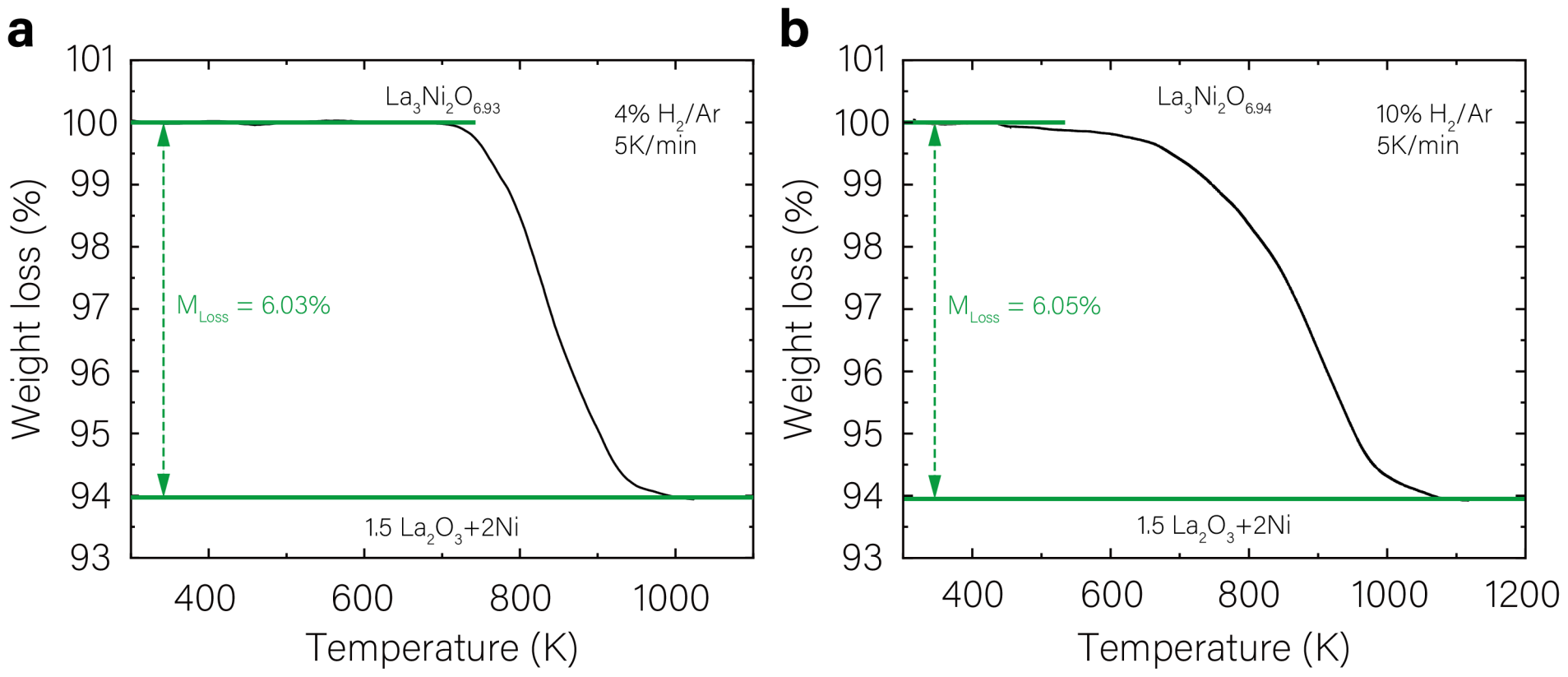

**Fig. S11 | Thermogravimetric analysis of $La_3Ni_2O_7$ crystals. a,b,** Thermogravimetric analysis (TGA) curves measured on two representative samples from the same growth batch as the crystals used in the neutron scattering experiments. The single crystals were crushed into small pieces, but not finely ground into powder; the typical particle size was on the order of 100 μm. The measurements were performed under reducing $H_2$/Ar mixed-gas atmospheres with a heating rate of 5 K $min^{-1}$. The broad reduction profiles likely reflect the relatively large particle size and diffusion-limited oxygen removal during reduction. The oxygen contents estimated from the total mass losses after reduction to 1.5 $La_2O_3$ + 2Ni are 6.93 and 6.94 per formula unit for the samples shown in **a** and **b**, respectively. These values indicate that the crystals are close to the nominal $La_3Ni_2O_7$ stoichiometry, with only a minor oxygen deficiency.

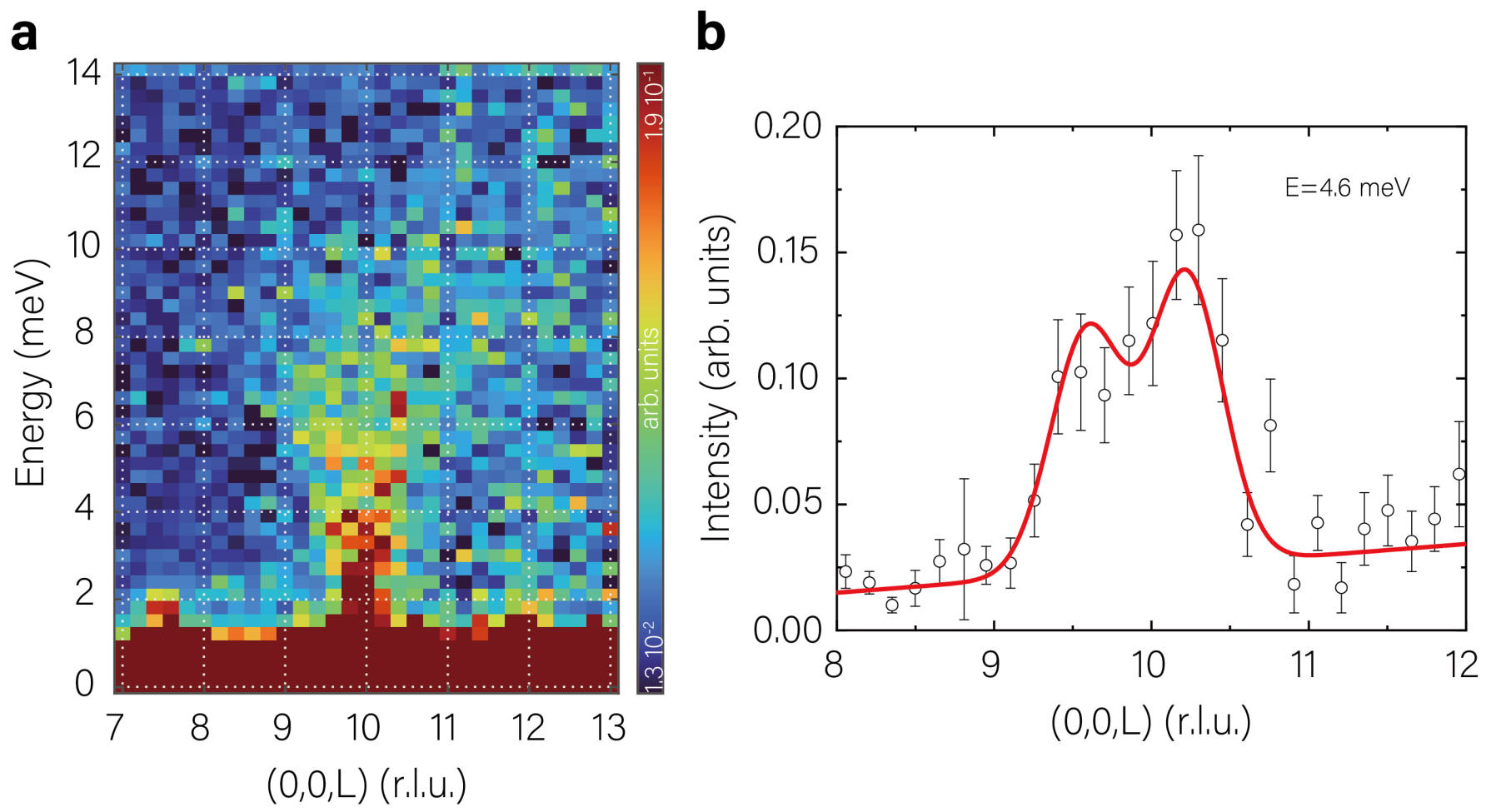

**Fig. S12 | Acoustic phonon used for absolute-intensity normalization. a,** Low-energy inelastic neutron scattering intensity map along the (0, 0, *L*) direction, showing an acoustic phonon branch near the nuclear Bragg position (0, 0, 10) at 163 K. **b,** Constant-energy cut at E = 4.6 meV along (0, 0, *L*), showing the momentum dependence of the acoustic phonon intensity. The cut was integrated over $H = -0.1$ to 0.1, $K = -0.1$ to 0.1, and $E = 4.6 \pm 0.5$ meV. The measured phonon intensity was compared with the calculated one-phonon neutron scattering cross section to determine the absolute scale factor used to convert the magnetic scattering intensity to $\chi''(Q,\omega)$ in absolute units.

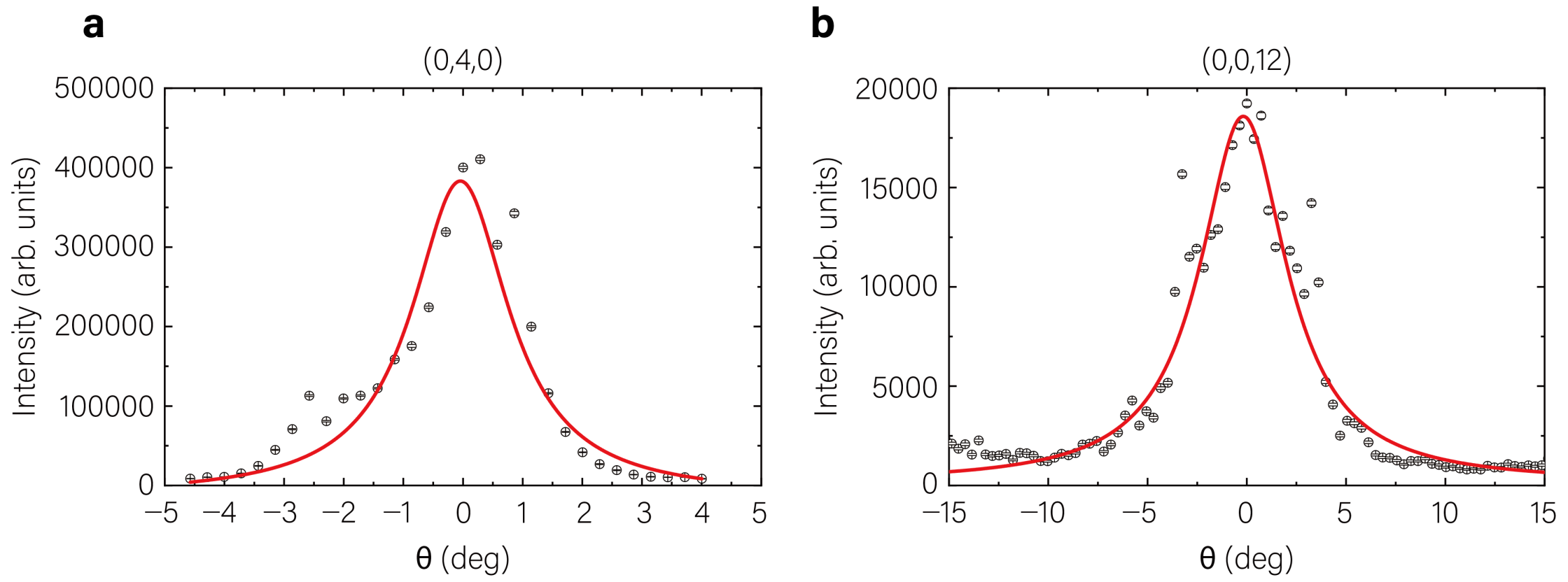

**Fig. S13 | Mosaic spread characterization of the co-aligned 25-crystal array. a,b,** Transverse scans through the nuclear Bragg reflections (0, 4, 0) and (0, 0, 12), respectively. The (0, 4, 0) scan gives an in-plane azimuthal mosaic width with FWHM = 1.98° (**a**), whereas the (0, 0, 12) scan gives a *c*-axis tilt distribution with FWHM = 5.30° (**b**).